\documentclass[11pt]{article}

\usepackage{amsmath,amssymb}
\usepackage{graphicx}
\usepackage{caption}
\usepackage{color}
\usepackage{dcolumn}
\usepackage{bm}
\usepackage{float}
\usepackage{epstopdf}
\usepackage[superscript]{cite}
\usepackage{xcolor}
\usepackage[nolist]{acronym}
\usepackage{braket}
\usepackage{blkarray}
\usepackage[textsize=tiny,backgroundcolor=yellow]{todonotes}

\newcommand{\rt}[1]{\textcolor{black}{#1}}   
\newcommand{\bt}[1]{\textcolor{black}{#1}}

\definecolor{background-color}{gray}{0.98}

\title{Theoretical X-Ray Spectroscopy of Transition Metal Compounds}

\author{Sergey I. Bokarev\thanks{Institut f\"ur Physik, Universit\"at Rostock, Albert-Einstein-Str. 23-24, 18059 Rostock, Germany}, Oliver K\"uhn\thanks{Institut f\"ur Physik, Universit\"at Rostock, Albert-Einstein-Str. 23-24, 18059 Rostock, Germany}}

\date{}
\begin{document}
\maketitle

\begin{acronym}[RASSCF  ]
\acro{ADC}{Algebraic Diagrammatic Construction}
\acro{AES}{Auger Electron Spectrum}
\acro{AMFI}{Atomic Mean-Field Integrals}
\acro{AO}{Atomic Orbital}
\acro{AS}{Active Space}
\acro{BO}{Born-Oppenheimer}
\acro{CAS}{Complete Active Space}
\acro{CASPT2}{Complete Active Space second order \ac{PT}}
\acro{CASSCF}{Complete Active Space \ac{SCF}}
\acro{CC}{Coupled Cluster}
\acro{CI}{Configuration Interaction}
\acro{CIS}{\ac{CI} Singles}
\acro{CISD}{\ac{CI} Singles Doubles}
\acro{CK}{Coster-Kronig}
\acro{CSF}{Configuration State Function}
\acro{CT}{Charge Transfer}
\acro{CVS}{Core-Valence Separation}
\acro{DCL}{Dynamical Classical Limit}
\acro{DFT}{Density Functional Theory}
\acro{DFT-CI}{Configuration Interaction on \ac{KS-DFT} orbitals}
\acro{DMRG}{Density Matrix Renormalization Group}
\acro{DO}{Dyson Orbital}
\acro{DOF}{Degree Of Freedom}
\acro{EOM}{Equation Of Motion}
\acro{EOM-CC}{Equation-Of-Motion Coupled Clusters}
\acro{FEL}{Free Electron Laser}
\acro{FOA}{Frozen Orbital Approximation}
\acro{FCI}{Full \ac{CI}}
\acro{HERFD-XAS}{High Energy Resolution Fluorescence Detection X-ray Absorption Spectrum}
\acro{GASSCF}{Generalized Active Space \ac{SCF}}
\acro{HF}{Hartree-Fock}
\acro{HHG}{High Harmonic Generation}
\acro{IOTC}{Infinite-Order Two-Component}
\acro{IP}{Ionization Potential}
\acro{ICD}{Interatomic Coulombic Decay}
\acro{KS-DFT}{Kohn-Sham Density Functional Theory}
\acro{LFM}{Ligand Field Multiplet}
\acro{LR}{Linear Response}
\acro{LR-CC}{Linear-Response \ac{CC}}
\acro{LR-TDDFT}{Linear-Response TDDFT}
\acro{MCSCF}{Multi-Configurational Self-Consistent Field}
\acro{MCTDHF}{Multi-Configurational Time-Dependent \ac{HF}}
\acro{MD}{Molecular Dynamics}
\acro{MO}{Molecular Orbital}
\acro{LCAO}{Linear Combinations of Atomic Orbitals}
\acro{MR}{Multi-Reference}
\acro{NEVPT2}{$N$-electron Valence second order \ac{PT}}
\acro{NXES}{Non-resonant X-ray Emission Spectrum} 
\acro{OEA}{One-Electron Approximation}
\acro{PES}{Photoelectron Spectrum}
\acro{PFY}{Partial Fluorescence Yield}
\acro{PP}{Polarization Propagator}
\acro{PT}{Perturbation Theory}
\acro{QC}{Quantum Chemistry}
\acro{RASPT2}{Restricted Active Space Perturbation Theory (second order)}
\acro{RASSCF}{Restricted Active Space Self-Consistent Field}
\acro{RASSI}{Restricted Active Space State Interaction}
\acro{REW}{Restricted Excitation Window}
\acro{REW-TDDFT}{Restricted Excitation Window Time-Dependent Density Functional Theory}
\acro{RIXS}{Resonant Inelastic X-ray Scattering}
\acro{ROHF}{Restricted Orbital HF}
\acro{ROCIS}{Restricted Open-shell Configuration Interaction Singles}
\acro{RPES}{Resonant Photoelectron Spectrum}
\acro{RPE}{Resonant Photoelectron}
\acro{RT}{Real-Time}
\acro{RXES}{Resonant X-ray Emission Spectroscopy}
\acro{SA}{Sudden Approximation}
\acro{SCF}{Self-Consistent Field}
\acro{SCL}{Static Classical Limit}
\acro{SOC}{Spin-Orbit Coupling}
\acro{SD}{Slater Determinant}
\acro{TCF}{Time-Correlation Function}
\acro{TD-CI}{Time-Dependent \ac{CI}}
\acro{TDDFT}{Time-Dependent Density Functional Theory}
\acro{TD-MCSCF}{Time-Dependent \ac{MCSCF}}
\acro{TISE}{Time-Independent Sch\"odinger Equation}
\acro{TD-RASCI}{Time-Dependent Restricted Active Space} 
\acro{TDSE}{Time-Dependent Sch\"odinger Equation}
\acro{TFY}{Total Fluorescence Yield}
\acro{TM}{Transition Metal}
\acro{TP-DFT}{Transition Potential DFT}
\acro{UPS}{Ultraviolet Photoelectron Spectrum}
\acro{UV/Vis}{UltraViolet/Visible photon energy range}
\acro{XAS}{X-ray Absorption Spectrum}
\acro{XC}{eXchange-Correlation}
\acro{XES}{X-ray Emission Spectrum}
\acro{XFEL}{X-ray Free Electron Laser}
\acro{XPS}{X-ray Photoelectron Spectrum}
\acro{XUV}{eXtreme UltraViolet}
\acro{ZORA}{Zero-Order Regular Approximation}
\end{acronym}

\begin{center}
\subsubsection*{\small Article Type:}
Advanced Review

\hfill \break
\thanks

\subsubsection*{Abstract}
\begin{flushleft}

\rt{X-ray spectroscopy is one of the most powerful tools to access structure and properties of matter in different states of aggregation as it allows to trace atomic and molecular energy levels in course of various physical and chemical processes.} X-ray spectroscopic techniques probe the local electronic structure of a particular atom in its environment, in contrast to UV/Vis spectroscopy, where transitions generally occur between delocalized molecular orbitals. Complementary information is provided by using a combination of different absorption, emission, \rt{scattering} as well as photo- and autoionization X-ray methods. However, interpretation of the complex experimental spectra and verification of experimental hypotheses is a non-trivial task and  powerful first principles theoretical approaches that allow for a systematic investigation of a broad class of systems are needed. 
Focussing on transition metal compounds, $L$-edge spectra are of particular relevance as they probe the frontier $d$-orbitals involved in metal-ligand bonding. Here, near-degeneracy effects in combination with spin-orbit coupling lead to a complicated  multiplet energy level structure, which poses a serious challenge to quantum chemical methods. \ac{MCSCF} theory has been shown to be capable of providing a rather detailed understanding of experimental X-ray spectroscopy. However, it cannot be considered as a 'blackbox' tool and its application requires not only a command  of formal theoretical aspects, but also a broad knowledge of already existing applications. Both aspects are covered in this overview.
\end{flushleft}
\end{center}

\clearpage

\renewcommand{\baselinestretch}{1.5}
\normalsize

\section{INTRODUCTION}  
%
Spectroscopy is one of the most powerful tools in physics and chemistry  
and, in particular, X-ray spectroscopy attracts special attention not at least due to the ongoing developments at 
\ac{XFEL} facilities and 
\ac{HHG} sources \cite{Chergui_SD_2016, Young_JPB_2018}.
As a consequence of the localized nature of the core orbitals, the transition operator acts locally and that is why X-ray excitation probes the local electronic structure of a particular atom embedded in its chemical environment.  
This is in contrast to, e.g., \ac{UV/Vis} spectroscopy, where transitions usually occur between delocalized valence \acp{MO}.  
The combination of different absorption, emission, \rt{scattering}, photoionization, and diffraction X-ray techniques allows addressing various aspects of static properties as well as photoinduced and chemical dynamics in steady-state and time-resolved X-ray spectroscopy \cite{Stoehr_book_1992, deGroot_Book, Milne_CCR_2014, Chergui_SD_2016, Chergui_CR_2017}. 
Sharpening the experimental probe does not guarantee, however, an increase of the acquired knowledge as the complexity of the detected signal increases as well when addressing more and more intricate effects. Thus, for interpretation of the experimental data, theoretical modeling is mandatory.
 
\bt{This review presents an overview on theoretical approaches with a focus on  first principles electronic structure methods and in particular on multi-reference wave function techniques.  
Applications to the interpretation of experimental data will be shown to provide a means for dissecting different structural and dynamical problems. We will restrict ourselves to isolated molecules and complexes as they appear in the gas or solution phases. Extended periodic systems like crystals are explicitly excluded from consideration as respective theoretical models represent a huge self-standing body of methods which are reviewed elsewhere.~\cite{Rehr_CCR_2005, Milne_CCR_2014, Leng_WIRCMS_2016}
Here, we address only discrete the pre-edge structure leaving XANES and EXAFS parts of the absorption edge aside since \rt{especially the latter} provides more information on the geometric rather than electronic structure.~\cite{Schnohr_XASoS_2015}}

\bt{Although computational X-ray methods have been developing apace with quantum chemistry since its infancy, the 21st century heralds an explosive growth of theoretical studies of X-ray spectra. These studies are mainly devoted to the $K$-edge ($1s$) spectra of second period and heavier elements using single reference methods. 
It is barely possible and not aimed in the current review to cover them all, including the  numerous applications (for reviews see References \citenum {Besley_PCCP_2010, DeBeer_CICI_2013, Milne_CCR_2014, Norman_CR_2018}).
Therefore, mostly the soft X-ray photon energy range (0.1-1.0 keV) and 4th period \ac{TM} complexes will be considered. 
Thus, mainly metal $L$-edges will be discussed corresponding to excitation or ionization from $2p$ core orbitals.} 

 %
\section{X-RAY SPECTROSCOPY}  
 
\subsection{ What makes it special?} 
%
Spectroscopy in the \ac{UV/Vis} spectral range is
a standard tool for probing electronic transitions and related
processes, taking place within the valence state manifold. Here, participating electronic states often exhibit rather delocalized electron densities. This makes an interpretation in terms of local changes difficult, e.g., with respect to specific bonds. 
In contrast, X-ray transitions involve at least one orbital, which is localized on an atom, thus providing a local probe of the electronic structure. 
Moreover, core level energies vary vastly between different
atoms what provides element specificity. 
Further, core orbitals with non-zero angular momentum lead to pronounced relativistic effects, e.g., caused by \ac{SOC}. Conventional X-ray spectroscopy is plagued by the rather short lifetime of the core-hole (4-10 fs), which limits the resolution (a possible way to overcome this problem is  \ac{HERFD-XAS} \cite{Hamalainen_PRL_1991}). 
However, this can be turned into an advantage by using the core-hole lifetime to clock ultrafast dynamics (``core-hole clock'') \cite{Piancastelli_JPB_2014}. Finally, X-ray wavelengths are in the range of relevant molecular length scales, enabling better spatial and time resolution.

The list of problems, which X-ray spectroscopy typically addresses includes probing the chemical environment and bond distances \cite{Pollock_JACS_2013, Schnohr_XASoS_2015, Soldatov_P_2018, Mastelaro_M_2018}, the nature of chemical bonds \cite{Suljoti_ACIE_2013, Engel_JPCB_2014, Kunnus_JPCB_2016} as well as oxidation, spin, and solvation states. \cite{Bokarev_JPCC_2015, Wernet_Nat_2015, Vanko_JPCC_2015, Kubin_CS_2018} Further, it is indispensable for the investigation of solids and surfaces.~\cite{Stoehr_book_1992,deGroot_Book,Huefner_Book, Sham_SRiMS_2018}
While steady-state X-ray spectroscopy is well-established, emerging time-resolved experiments are pushing the limits to enter the regime of even the fastest nuclear and electron dynamics. This becomes possible due to the fact that the optical cycle of X-rays is in the few attosecond range.
Among others \cite{Chergui_SD_2016, Chergui_CR_2017}, recent examples include, e.g., the study of the (photo)reaction dynamics of acetylene isomerization upon core-hole formation \cite{Liekhus-Schmaltz_NC_2015}, photodissociation of simple flourides \cite{Pertot_S_2017}, ligand exchange \cite{Wernet_Nat_2015} and transient photodynamics \cite{Hong_ACR_2015, Worner_SD_2017, Jay_JPCL_2018} in metal complexes, photocatalytic processes on metal surfaces \cite{Newton_C_2017, Beye_JPCL_2016}, nanosecond dynamics of dissociation and reassociation of insulin \cite{Rimmerman_JPCL_2017}, and ultrafast electron and nuclear wave packet dynamics using time-resolved and transient photoelectron spectroscopy \cite{Brausse_PRA_2018a, Squibb_NC_2018}. 

The advantages of X-ray as compared to \ac{UV/Vis} spectroscopy come at the price of requiring large scale  facilities like synchrotrons and \acp{XFEL}. 
However, recent progress in \ac{HHG} sources scales down the size of the setups to the table-top. 
\rt{Moreover, prominent recent developments include low-power X-ray (bremsstrahlung) tubes~\cite{Seidler_RSI_2014} and plasma-based  X-ray pulse generation~\cite{Mantouvalou_RSI_2015}. }
It is anticipated that within a few years a table-top intense and stable source of isolated ultrafast XUV and X-ray pulses will appear what will herald the ``golden age'' of frequency- and time-resolved X-ray spectroscopy giving rise to a plethora of new techniques and perspective directions \cite{Young_JPB_2018}.

\subsection{Types of X-ray probes} 
 
\begin{figure}[bth] 
	\centering 
	\includegraphics[width=0.95\textwidth]{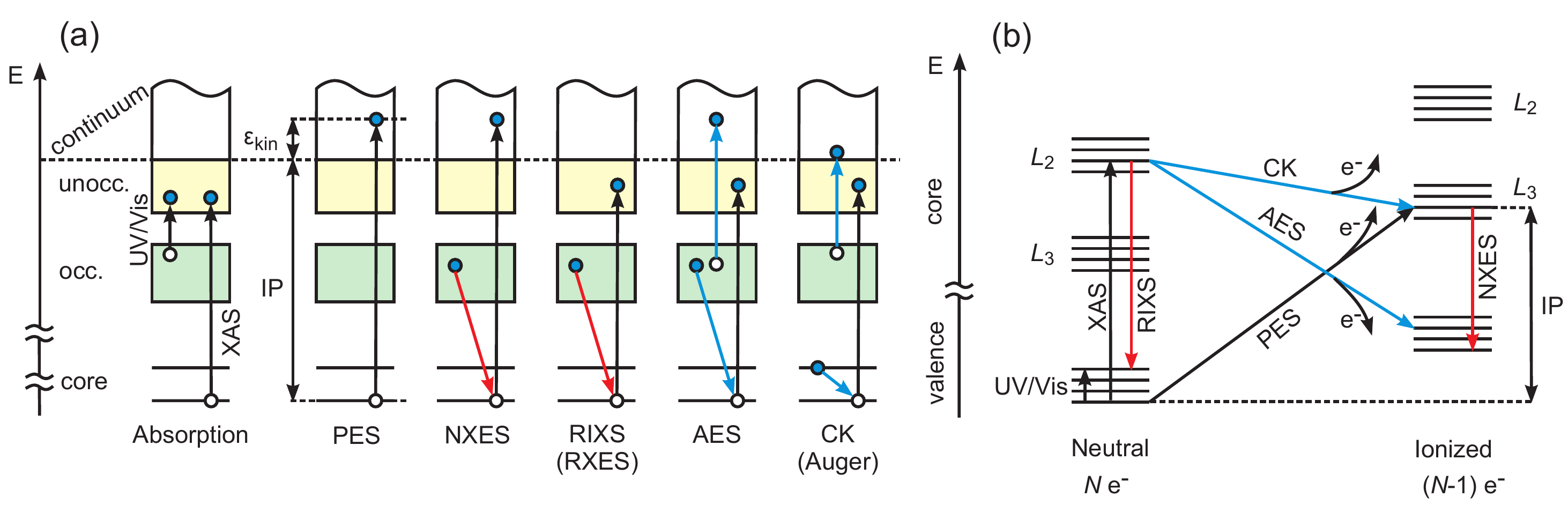}
	\caption{\label{fig:orb_state_picture} 
		 (a) Processes relevant for X-ray spectroscopy viewed from the \ac{MO} picture standpoint. For absorption, the respective process in \ac{UV/Vis} range is also shown. 		 (b) Processes relevant for X-ray spectroscopy in the many-body state picture. (Color code for arrows: photon absorption  --  black, radiative decay of a core hole -- red, non-radiative decay -- blue). 
		 \rt{Abbreviations: \acf{PES}, \acf{NXES}, \acf{RIXS}, \acf{RXES}, \acf{AES}, \acf{CK}.}
               }	 
\end{figure} 
For \ac{TM} compounds, $L$-edge spectra  enjoy great popularity~\cite{deGroot_Book, Milne_CCR_2014}. 
They are  due to excitation or ionization from the $2s$ and $2p$  orbitals ($L$-shell) of the target atom to bound or continuum states, respectively. Being interested in frontier $d$-orbitals containing information on the nature of metal-ligand or metal ion-solvent interaction, dipole-allowed excitation from $2p$ orbitals is addressed more frequently.  Traditionally, solid state samples have been investigated. However, recent developments in the vacuum liquid microjet technique have broadened the applicability of soft X-ray spectroscopy to highly volatile liquid solutions.~\cite{Winter_CR_2006, Smith_CR_2017}
 
Excitation of a system from its electronic ground state to a high-energy and highly non-equilibrium core-excited state triggers a number of relaxation processes which either represent losses or can be used as a spectroscopic probe of material properties. 
Let us consider them on the example of $2p$ $L$-edge spectroscopy as illustrated in Figure~\ref{fig:orb_state_picture}. In general there are two pictures with respect to the electronic states that can be used for the discussion. First, the intuitive and therefore widely used ``single electron''  \ac{MO} picture (panel (a)). Second, the more appropriate ``state picture'', which is based in the many-particle eigenstates of the electronic Hamiltonian (panel (b)).

Upon absorption of a soft X-ray photon (0.1-1.0 keV), a $2p$ core hole is created at the metal center accompanied by   $K$-edge ionization of the ligands. 
This process is depicted on the very left of Figure~\ref{fig:orb_state_picture} and is   subject of  X-ray absorption spectroscopy \cite{Schnohr_XASoS_2015}.  
In this case, an electron is excited from a localized core \ac{MO} to delocalized bound one. Within the state picture, the \ac{XAS} is due to transitions within the $N$-electron manifold, i.e.\ between the ground  and core-excited states of the (``neutral'') system.

The core-electron can also be excited into the continuum and the photocurrent of the outgoing electrons can be measured yielding a \ac{PES}. In the state picture this corresponds to a transition between the initial $N$-electron (``neutral'') system and the $N-1$-electron ionized system as illustrated in Figure~\ref{fig:orb_state_picture}b. The proper total final state is formed by the antisymmetrized product of the   state of the $N-1$ electron manifold and the outgoing free electron state.

The created core-hole can decay either radiatively or non-radiatively. Radiative decay channels comprise $3d \rightarrow 2p$ and $3s\rightarrow 2p$ pathways, which represent about 1\% of the total decay probability for the first-row \acp{TM} \cite{Kotani_RMP_2001}.  
Depending on whether the initial excitation has been into a bound or continuum state one speaks about resonant (RXES)  or \ac{RIXS} \cite{Gelmukhanov_PR_1999} or the \ac{NXES}, respectively. 
The major decay channel is due to Auger, \ac{CK} decay, and \ac{ICD} autoionization processes \cite{Averbukh_JESRP_2011}  (Figure~\ref{fig:orb_state_picture}a). 
In the state picture these processes involve a transition between the $N$ and $N-1$-electron manifolds and are reflected in the \ac{AES}.

In general \ac{XAS} and  \ac{RIXS} are called photon-in/photon-out spectroscopies, whereas \ac{PES} and \ac{AES} are photon-in/electron-out techniques.
The combination of X-ray absorption and emission together with photoelectron studies  provides a very powerful suite of tools to address electronic structure and specific solute-solvent interactions \cite{Milne_CCR_2014,Golnak_SR_2016}. 
%
\subsubsection{Absorption spectroscopy} 
Absorption spectroscopy provides the most simple type of spectra considered in the context of X-ray excitation as long as the discrete pre-edge structure is considered only.  Viewed in the \ac{MO} picture, \ac{XAS} probes the unoccupied valence orbitals.
Conventional X-ray experiments operate in the  weak-field regime \cite{Schnohr_XASoS_2015} and the respective absorption amplitude, proportional to the absorption cross section at frequency $\Omega$, can be expressed in terms of first-order time-dependent perturbation theory (Fermi's Golden Rule) \cite{Ballentine_Book}:

\begin{equation}\label{eq:XAS} 
{\mathcal{X}}(\Omega) = \sum\limits_{g}{f(E_{g},T)\sum\limits_{i}^{}{\left| \bra{i} \hat{\mathbf{d}}  \cdot \mathbf{e}_{in}\ket{g}\right|_{}^{2}}\Lambda(E_{g}^{}+\Omega-E_{i}^{})} . 
\end{equation} 
Here, the $\ket{g}$ and $\ket{i}$ states are ground and core-excited final states with respective energies $E_g$ and $E_i$. 
Due to finite temperature $T$, there can be several initial states populated according to the Boltzmann distribution $f(E_g,T)=\exp{(-E_g/kT)}/\sum_j \exp{(-E_j/kT)}$.  This especially applies to \ac{TM} complexes, where the initial high degeneracy of the electronic states can be lifted due to Jahn-Teller effect and \ac{SOC}, leading to several levels accessible by thermal excitation. 

Further, $\hat{\mathbf{d}}$ and $\mathbf{e}_{in}$ in Eq.~\ref{eq:XAS} are electric dipole operator and polarization vector of the incoming light.
The lineshape function, $\Lambda(E_{g}^{}+\Omega-E_{i}^{})$, characterizes the density of the final states and, thus, the form of the absorption bands. Effects to be considered here include the width of the incoming excitation pulse, homogeneous (lifetime) broadening due to Auger and radiative decay, and inhomogeneous broadening due to influence of environment, e.g., solvent or phonon broadening in solid state.\cite{May_book_2011} These broadening parameters can be often fitted to best reproduce the experiment. 

To address \ac{XAS} theoretically in dipole approximation, one thus needs to calculate energies of the core-excited electronic states with respect to the ground one and the ground-to-core-excited transition dipole moments. Note that for hard X-ray radiation the dipole approximation breaks down and higher order multipole moments have to be considered.\cite{Bernadotte_JCP_2012}
 
For $L$-edge spectra,  additional complexity enters due to  strong \ac{SOC}, triggered upon core-hole formation and within the $3d$ states themselves \cite{deGroot_Book}.  This leads to a characteristic shape of the spectrum containing two groups of bands, so-called $L_3$ and $L_2$ edges (see Figure~\ref{fig:XAS_RIXS} below). 
%
\subsubsection{Resonant emission spectroscopy (\ac{RIXS})} 
%
Staying in the weak-field perturbational regime, the radiative decay of the core-excited state $\ket{i}$ prepared upon absorption of X-ray light can be described within second-order perturbation theory.  The respective Kramers-Heisenberg expression for the emission amplitude reads \cite{Ballentine_Book}
\begin{equation}\label{eq:RIXS} 
\begin{split} 
{\mathcal{R}}(\Omega,\omega) &= \sum\limits_{g}{f(E_{g},T)\sum\limits_{f}^{}{\left|\sum\limits_{i}^{}{\sqrt{\frac{\Gamma_{i}}{\pi}}\frac{\bra{f} \mathbf{e}_{\rm out}^\ast \cdot \hat{\mathbf{d}}\ket{i}\bra{i} \hat{\mathbf{d}}  \cdot \mathbf{e}_{\rm in}\ket{g}}{E_{g}^{}+\Omega-E_{i}^{}-i\Gamma_{i}^{}}}\right|_{}^{2}}} \\ 
&\times{\Lambda}(E_{g}^{}+\Omega-E_{f}^{}-\omega). 
\end{split} 
\end{equation} 
It contains the same ingredients as the XAS expression, Equation~\eqref{eq:XAS}, where in addition $\omega$ and $\mathbf{e}_{out}$ denote the energy and polarization of the emitted photon, $\Gamma_i=1/\tau_i$ is the inverse lifetime of core-excited state $\ket{i}$ and state $\ket{f}$ represents the final valence state. Notice that due to the fact that the sum over $i$ appears under the square, the radiative channels ending up in the same final state $f$ interfere with each other. 
This photon-in/photon-out process represents an electronic analogue of the well-known vibrational Raman spectroscopy. 
The only difference is that one addresses transitions not between  vibrational levels but between different electronic states. 
Formally, RIXS corresponds to a non-linear $\chi^{(3)}$ process and therefore is analogous to four-wave mixing but with the last two interactions being of spontaneous rather than stimulated nature \cite{Mukamel_book_1999}.  Thus, it is only linearly proportional to the intensity of the incoming light. Drawing on this analogy the overall spontaneous light emission signal can be separated into two contributions: sequential incoherent fluorescence where a population in the excited state is created followed by radiative decay and a direct ``coherent'' Raman process. While in the former case resonant excitation is required, RIXS also works for off-resonant excitation, i.e. involving only coherences between the participating states.

 In the following we will focus on RIXS only. 
The RIXS intensity can be recorded as a function of incoming, $\Omega$, and outgoing, $\omega$, photon energy. Thus, it provides a two-dimensional spectrum, resolving both absorption and emission processes. Often one discusses one dimensional cuts at fixed excitation energy of this spectrum only. 
Alternatively, the so-called \ac{PFY} spectrum at a given excitation energy $\Omega$ can be obtained by integrating with respect to the emission energies in some energy window $\omega \in [\omega_0,\omega_1]$
\begin{equation}\label{eq:PFY}
\mathcal{F}(\Omega)=\int_{\omega_0}^{\omega_1}\mathcal{R}(\Omega,\omega)d\omega \, ,
\end{equation}
 providing a spectrum analogous to \ac{XAS}. Viewed in the orbital picture (Figure \ref{fig:orb_state_picture}a), RIXS gives access to the unoccupied \acp{MO} as well as to the occupied valence  \acp{MO}.  In state picture (Figure \ref{fig:orb_state_picture}b), absorption and emission processes become coupled because of orbital relaxation due to core-hole and valence excitation relaxation. Thus,  \ac{RIXS} approximately maps the valence excited states. Calculating \ac{RIXS} is a  more elaborate task as compared to \ac{XAS} since these core-hole and valence excitation states may have different requirements to the computational protocol. 
 %
 %
\subsubsection{Photon-in/electron-out spectroscopy} 
Photon-in/electron-out spectroscopy represents a powerful suit of methods 
\cite{Huefner_Book, Fadley_JESRP_2010, Fadley_JESRP_2014}. 
Depending on whether the X-ray photon energy exceeds the binding energy of the core-electron or is below the core ionization threshold one distinguishes between non-resonant PES and  \ac{RPES}, cf.  Figure~\ref{fig:orb_state_picture}a.
\ac{PES} provides direct information on the energies of core orbitals, e.g., the value of their \ac{SOC} splitting, if viewed from the orbital picture perspective. In the state picture, it corresponds to a transition from e.g. ground state of non-ionized system to the core-excited states of ionized one. 
Moreover, the shake-up excitations, where ionization is accompanied by valence excitation, can provide valuable insights into the electronic and geometric structure.~\cite{Norell_PCCP_2018a}
 
The resonant counterpart, \ac{RPES},  contains two processes -- direct ionization from the valence level and autoionization of the core-excited state. 
The former one represents a usual \ac{PES} where electrons from valence \acp{MO} are ejected having  kinetic energies comparable to that of the absorbed X-ray photon. 
The latter one is an Auger process, i.e. a transition between core-excited state of the non-ionized manifold and the valence state of the ionized manifold accompanied by emission of an electron as depicted in Figure~\ref{fig:orb_state_picture}b. 
In terms of orbital picture, it corresponds to core-hole refill  by a valence electron and simultaneous emission of another valence electron with high kinetic energy.

The total \ac{RPES}  cross section assuming integration over all directions of the outgoing photoelectron reads 
\begin{equation}\label{eq:RPES} 
\begin{split} 
{\mathcal{P}}(\Omega,\varepsilon_{\rm kin})&=\sum_i{f(E_g,T)}\sum_{f^+} \Lambda(E_{f^+}+\varepsilon_{\rm kin}-E_g-\Omega) \\ 
&\times\left|\bra{f^{+}\psi^{\rm el}(\varepsilon_{\rm kin})}\mathbf{e}_{\rm in}\cdot\hat{\mathbf{d}}\ket{g}+ 
\sum_i\frac{\bra{f^{+}\psi^{\rm el}(\varepsilon_{\rm kin})} \hat{H}-E_i\ket{i}\bra{i} \hat{\mathbf{d}}  \cdot \mathbf{e}_{\rm in}\ket{g}}{E_{g}^{}+\Omega-E_{i}^{}-i\Gamma_{i}^{}}\right|^2 \enspace . 
\end{split} 
\end{equation}
The structure of this expression is a hybrid of first-order Fermi's Golden rule for direct photoionization (first term under square) and second order Kramers-Heisenberg-like term for autoionization. 
The system starts from thermally populated initial states $\ket{g}$. 
The incoming X-ray light either ionizes the system by transition to a final state $\ket{f^{+}\psi^{\rm el}(\varepsilon_{\rm kin})}$, which is the properly antisymmetrized product of the bound $N-1$-electron ionic remainder state $\ket{f^+}$ and the state of the free electron $\ket{\psi^{\rm el}(\varepsilon_{\rm kin})}$, having kinetic energy $\varepsilon_{\rm kin}$. 
Alternatively, the same state can be obtained by excitation to the intermediate bound core-excited state $\ket{i}$, which then decays non-radiatively, mediated by the Hamiltonian $\hat{H}$ of the system.  This decay is essentially due to the Coulomb coupling which is dominating in the respective matrix element. 
 
There are some more comments needed at this point. 
First of all, due to the coinciding final state of both processes there is an interference between the direct and Auger parts of the signal. 
This interference is present also for decay channels from different intermediate states in the Auger part of the signal.
Second, Equation \eqref{eq:RPES} sheds light on the nature of the parameters $\Gamma_i$  entering also the \ac{RIXS} expression, Equation~\eqref{eq:RIXS}. 
In fact, $\Gamma_i=2\pi\sum_{f^+}|\bra{f^{+}\psi^{\rm el}(\varepsilon_{\rm kin})} \hat{H}-E_i\ket{i}|^2$, representing the total Auger decay rate of core-excited state $\ket{i}$. 
In principle, radiative decay would add to the total rate, but for TMs autoionization is the dominating process. 
Next, we note that here we assumed the simple product form $\ket{f^{+}\psi^{\rm el}(\varepsilon_{\rm kin})}$ of the final state and thus correlations between the ionic remainder and the  outgoing electron are neglected. 
A more elaborate scheme taking interaction of ionization channels requires a scattering theory formulation \cite{Cortes_JPB_1994}.
Finally,  \ac{RPES} can be also viewed as a 2D type of spectrum \cite{Golnak_SR_2016} as it depends on the incoming photon energy $\Omega$ and the photoelectron kinetic energy $\varepsilon_{\rm kin}$.  
 
As compared to \ac{XAS} and \ac{RIXS}, \ac{PES} is analogous to the former and \ac{AES} to the latter. 
The difference is that \ac{PES} contains information on the core-excited states of the ionic remainder, whereas \ac{AES} contains that of the core states of the non-ionized system and valence manifold of ionized system (see Figure~\ref{fig:orb_state_picture}b). 
\ac{RPES} calculations are more elaborate due to the need to compute integrals between bound and continuum states to obtain matrix elements of the of the dipole moment and molecular Hamiltonian $\hat{H}-E_i$.  
 
In the weak field regime, \ac{PES} cross sections can be calculated in the framework of the \ac{DO} formalism \cite{Grell_JCP_2015, Oana_JCP_2007, Mohle_JCTC_2018}.  
For illustration, we leave the second term in Equation~\eqref{eq:RPES} aside and consider  only the direct photoionization represented by the first term. 
Assuming the strong orthogonality between the state $\ket{\psi^\mathrm{el}(\varepsilon_{\rm kin})}$ and \acp{MO} of the initial system, the expression can be rewritten in terms of  quasi-one-electron states $|\Phi^\mathrm{DO}_{f^+g}\rangle$ which are called \ac{DO}s. 
It represents an $N-1$-electron integral over initial and final ionic states $|\Phi^\mathrm{DO}_{f^+g}\rangle=\sqrt{N}\langle f^+ | g \rangle_{N-1}$. 
Using the \ac{DO}, the transition matrix element (neglecting polarization dependence) can be expressed as  
\begin{equation}\label{eq:DO_ME} 
   \bra{f^{+}\psi^{\rm el}(\varepsilon_{\rm kin})} \hat{\mathbf{d}} \ket{g}_N  
  \approx 
 \bra{\psi^{\rm el}(\varepsilon_{\rm kin})} \hat{\mathbf{d}} \ket{\Phi^{\rm DO}_{f^+g}}_1 \enspace .  
\end{equation} 
Here, the subscripts $N$ and 1 denote $N$-electron and one-electron integration, respectively. 
This   formal reorganization allows to separate system and electronic structure method dependence of the intensity from the actual representation of the photoelectron. 
Once the \ac{DO} is computed for a particular pair of bound states, $\ket{g}$ and $\ket{f^+}$, it can be further used to estimate transition dipoles for different kinetic energies $\varepsilon_{\rm kin}$ and even different forms of $\ket{\psi^{\rm el}}$. Commonly used  free electron functions include the plane wave expansion in terms of partial waves or Coulomb waves \cite{Oana_JCP_2007}. 
The former is appropriate when photodetachment from a negatively charged species is considered, whereas the latter accounts for spherical Coulomb potential of the ionic remainder. In case of  TM complexes, one needs many terms in the expansion to realistically approximate the free electron function \cite{Grell_JCP_2015}. 
This approach has facilitated interpretation of experimental XUV and X-ray \acp{PES} of transition metal complexes in solution.~\cite{Golnak_SR_2016, Moguilevski_CPC_2017, Raheem_SD_2017, Engel_PCCP_2017} 
 
Finally, we would like to mention the so-called \ac{SA}~\cite{Aberg_PR_1967}. It assumes that  the cross section for a pair of states, $\ket{i}$ and $\ket{f}$, is proportional to   $|\Phi^{\rm DO}_{fi}|^2$.  Thereby the kinetic energy dependence of the cross section is neglected such that no bound-to-continuum integrals in Equation~\eqref{eq:DO_ME} need to be calculated.
This approximation is usually considered to be justified if the nature of the transitions and thus of the  \acp{DO} is similar and kinetic energies are large.  
%
\section{THEORETICAL METHODS}

\begin{figure}[tbh]
	\centering
	\includegraphics[width=0.9\textwidth]{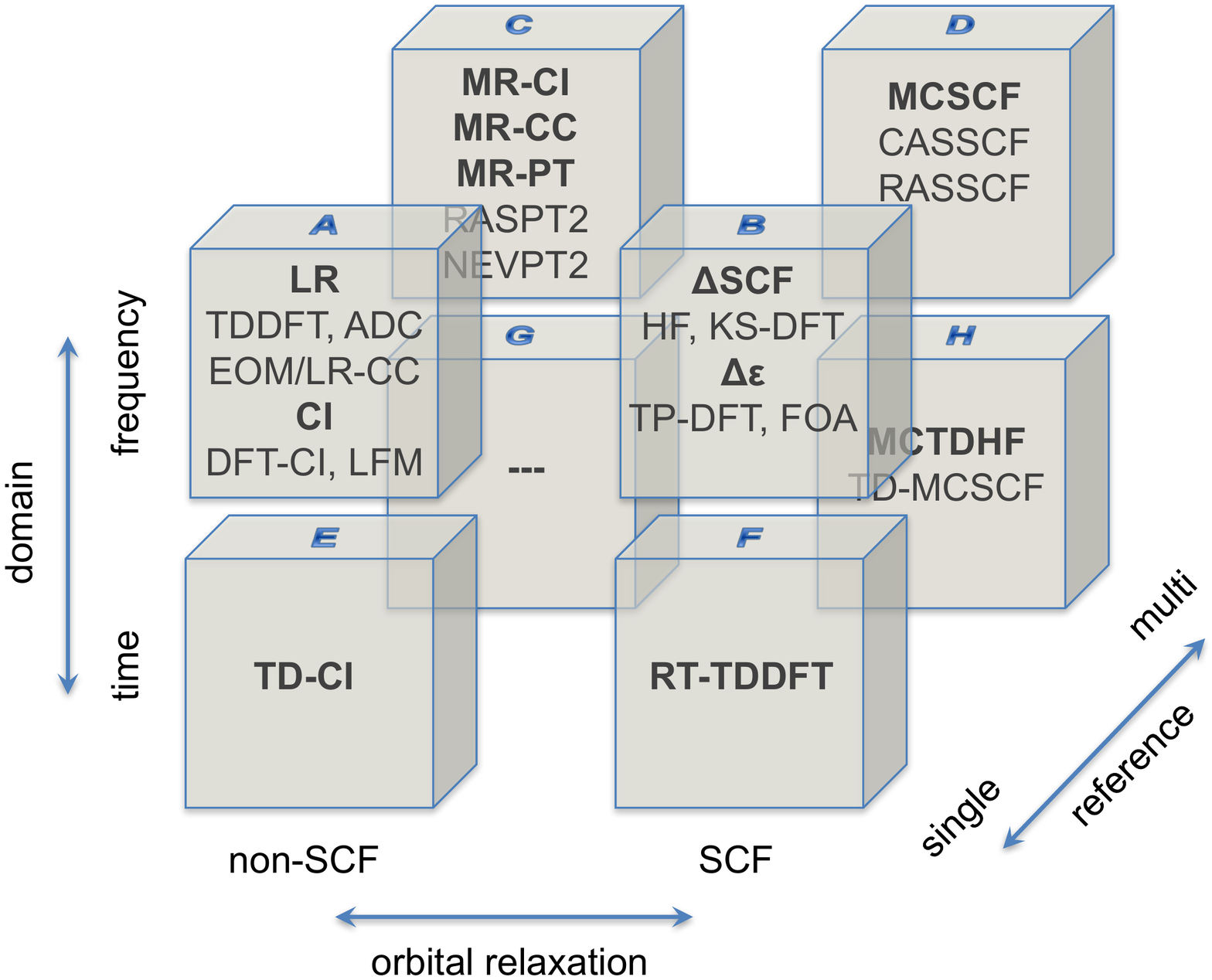}
	\caption{\label{fig:overview}
		 Overview on different methods discussed in this review, grouped according to three criteria: (i) frequency (energy) vs. time domain, (ii) explicit orbital optimization, i.e. \ac{SCF} vs. non-\ac{SCF}, and (iii) reference for construction of excited state basis is based on a single or multiple electronic configurations. 
		 (MR-CI, MR-CC, and MR-PT are multi-reference \ac{CI}, \ac{CC}, and \ac{PT}, respectively. Other abbreviations are defined in the text.) 
            }
\end{figure}
%

\subsection{General remarks}
%
Theoretical studies of X-ray spectra can be divided into two classes: 1) investigations of closed-shell molecules containing mainly the second-period atoms addressed with single-reference methods and 2) works reporting on open-shell systems, primarily \ac{TM} complexes. In general, first principles quantum chemical predictions for excited electronic states of \ac{TM} compounds are significantly more difficult than for compounds containing main group elements.~\cite{Neese_CCR_2007} 
\ac{TM} chemistry is closely connected to near degeneracy effects, also called static correlation, making methods based on a single \ac{SCF}-like reference configurations unreliable. Thus, computationally more demanding  multi-reference approaches are often required to obtain quantitative accuracy. 
In addition, the treatment of $d$-electrons requires the inclusion of  dynamic electron correlation. Finally, especially for core-excited states, relativistic effects, e.g., \ac{SOC}, have to be taken into account.

 An overview of available methods is provided in Figure~\ref{fig:overview}. In general, corresponding methods can be formulated in time (E,F,G,H) and energy (A,B,C,D) domain, which amounts to solving either the time-dependent or stationary  Schr\"odinger equation.
The spectroscopic observable (Equations~\eqref{eq:XAS},~\eqref{eq:RIXS}, and~\eqref{eq:RPES}) are formulated in terms of the eigenvalues and eigenstates of the stationary Schr\"odinger equation, $\hat{H}\ket{\Psi}=E \ket{\Psi}$.
However, the spectra could be equivalently expressed in as Fourier transforms of  multi-time dipole correlation functions.\cite{Ballentine_Book} For instance for \ac{XAS}, these are two-point and for \ac{RIXS} four-point dipole correlation functions \cite{Karsten_JPCL_2017, Karsten_JCP_2017, Karsten_JCP_2018, Karsten_JCP_2018a}.
Beyond this perturbational regime, i.e. for strong field excitation, or in cases of (shaped) ultrashort laser pulses one has to resort to explicit solution of the time-dependent Schr\"odinger equation including the field-matter interaction. Compared to the \ac{QC} methods,  \ac{RT} propagation approaches have received less attention so far.

Another, ``dimension'' of the classification in Figure~\ref{fig:overview} is with respect to the type of \acp{MO} that are used to express the many-electron states. In fact, every calculation starts from an \ac{MO} optimization; the question is whether it is done within the actual excited state method as well (B,D,F,H) or the \acp{MO} are fixed during excited state calculations \bt{(A,C,E)}.
Finally, \ac{QC} methods can be distinguished according to the used reference space into single (A,B,E,F) and multi-configuration \bt{(C,D,H)} approaches.
\bt{Note that we are not aware of the methods which fall into category G.}

In the following, we will review ab initio \ac{QC} methods, focussing on \ac{MCSCF}-based methods in frequency domain (Figure ~\ref{fig:overview}D and C) as applied to X-ray spectroscopy of \ac{TM} complexes. However, other methods mentioned in Figure~\ref{fig:overview} will be briefly discussed as well. 

\subsection{\bt{Basic concepts: \ac{SCF}}}
\subsubsection{\ac{HF} theory}
The \ac{HF} method is the most basic \ac{MO} approach discussed here. Although it is not directly applied to core-level spectroscopies it is central for understanding other methods.
It can be characterized by the following points:
(i) The $N$-electron wave function $\Psi^N(\mathbf{r}_1,...,\mathbf{r}_N)$, describing the ground state of the system, is represented as a single Slater determinant
$\Phi^N(\mathbf{r}_1,...,\mathbf{r}_N)=\det\{\phi_a(\mathbf{r}_1)\phi_b(\mathbf{r}_2)...\phi_c(\mathbf{r}_N)\}$ of  one-electron \acp{MO} $\{\phi_i(\mathbf{r}_j)\}$  (for inclusion of spin, see Section \ref{sec:Relativism}).
(ii) The optimal set of $\{\phi_i(\mathbf{r}_j)\}$ is searched variationally to obtain $\ket{\Phi^N}=\ket{\Phi_g}$ giving the best  possible estimate $E^{\rm HF}$ of the exact ground state energy $\bar E$ under the restriction of (i), see Eq.~\eqref{eq:HF}. 
(iii)    The electronic Hamiltonian contains one- and two-electron terms $\hat{H}=\sum_i \hat{h}(i) + \sum_{ij} \hat{g}(i,j)$. \ac{HF} theory is a  mean-field approach where the  latter are approximated by a one-electron potential $\sum_{ij} \hat{g}(i,j)\approx\sum_i \hat{V}^{\rm eff}(i)$.
Matrix elements of this one-electron operator are expressed via Coulomb, $J_{kl}$, and exchange, $K_{kl}$, two-electron integrals as $\bra{\phi_k}\hat{V}^{\rm eff}(1)\ket{\phi_k}=\sum_l (J_{kl}-K_{kl})$.
(iv) The problem is reformulated in terms of a   one-electron eigenvalue equation $\hat{f}_i \ket{\phi_i} = \varepsilon_i \ket{\phi_i}$, yielding orbitals $\phi_i(\mathbf{r}_j)$ and their energies $\varepsilon_i$. Since the Fock operator, $\hat{f}_i$,  depends on the orbitals $\{\phi_i(\mathbf{r}_j)\}$ the equation is solved iteratively (\ac{SCF} scheme). Essentially,  this is done via a search for unitary transformations $\hat U(\hat{\kappa})=\exp(\hat\kappa)$ of orbitals which give a minimal ground state energy 
\begin{equation}\label{eq:HF}
E^{\rm HF}(\pmb{\kappa})=\min_{\pmb{\kappa}} \bra{\Phi_g}\exp(-\hat\kappa)\hat H \exp(\hat\kappa)\ket{\Phi_g}	 \,
\end{equation}
where $\hat \kappa$ is an anti-hermitian one-electron operator. Thus, the orbital variations  are parametrized in terms of the elements of the skew-hermitan matrix $\pmb{\kappa}$, which
represents   $\hat\kappa$ in the basis of \acp{MO}; one may think of the elements of $\pmb{\kappa}$ as corresponding to pairwise orbital rotations.

In practice, one uses  the \ac{LCAO} form, where the \acp{MO} are represented in an atomic basis, $\{\chi_k(\mathbf{r} )\}$, 
\begin{equation}
\phi_i(\mathbf{r} )=\sum_k c_{ik}\chi_k(\mathbf{r} )	\, .
\end{equation}
Thus the  coefficients $\{c_{ik}\}$ are variationally optimized.

According to its mean-field nature, \ac{HF} theory does not take into account correlations due to the electron-electron interaction. Electron correlation is at least a two-body effect and should be distinguished from the one-electron orbital relaxation accounted for via the \ac{SCF} procedure. Different types of correlations can be identified if we adopt for the moment a classical point of view. 
Two electrons in an atom will try to ``occupy'' orbits with different radii (radial correlation) or reside on different sides of the nucleus (angular correlation).
This type of correlation is called dynamic (or weak) and can be accounted for by allowing $\ket{\Psi^N}$ to have in addition to the leading \ac{HF} term $\ket{\Phi^N}$ a small admixture of other determinants built on the same set $\{\phi_i\}$ but with different integer occupation numbers (excited determinants). 
This approach will be considered later in the context of the \ac{CI} method in Section \ref{sec:CI}.
There are cases, when a single leading term for $\ket{\Psi^N}$ cannot be found since several of determinants have comparable weights.
This more extreme case leads to complete break-down of \ac{HF} approximation and is referred to as static (or strong) correlation. It will be considered in the framework of \ac{MCSCF} theory in Section \ref{sec:MCSCF}.
Finally, one may distinguish differential correlation, when the correlation energy is different in different electronic states.

%
\subsubsection{\ac{KS-DFT}}
The \ac{KS-DFT} method belongs to the same class as \ac{HF}, see Figure~\ref{fig:overview}B. The approach is rather different though. Backed by the  Hohenberg-Kohn theorems it formulates the energy as a functional of electron density $\rho(\mathbf{r})$ subject to the variational principle.
Since $\rho(\mathbf{r})$ is a three-dimensional quantity and thus simpler than the $3N$-dimensional wave function, it offers distinct benefits from the viewpoint of computational efficiency.
In practice, one considers the Kohn-Sham scheme, where a single determinant is postulated for the ground state. The density is determined as $\rho^{\rm KS}(\mathbf{r})=\sum_{i} |\phi^{\rm KS}_i(\mathbf{r})|^2$, with  $\{\phi^{\rm KS}_i(\mathbf{r})\}$ being  the occupied Kohn-Sham \acp{MO}. 
This approach aims at the exact description of a system, adopting a formally non-interacting electron ansatz for the wave function. The ``non-classical'' terms which constitute the difference between this \ac{HF}-like solution and the exact one are thus transferred to the one-electron potential $\hat V^{\rm eff}(i)$. 
The energy functional of the electron density which is variationally minimized reads $E[\rho^{\rm KS}(\mathbf{r})]=T^{\rm KS}[\rho^{\rm KS}(\mathbf{r})]+J[\rho^{\rm KS}(\mathbf{r})]+E_{\rm el-nuc}[\rho^{\rm KS}(\mathbf{r})]+E^{\rm XC}[\rho^{\rm KS}(\mathbf{r})]$, containing kinetic energy of non-interacting system, ``classical'' Coulomb repulsion energy, electron-nuclear attraction energy, and the \ac{XC} energy $E^{\rm XC}[\rho^{\rm KS}(\mathbf{r})]$   comprising the difference between exact and approximate ($T^{\rm KS}$) kinetic energies, the effect of electron correlation, and exchange interaction. 
Various approximate $E^{\rm XC}$ exist and numerous techniques for the construction of this functional have been devised. 

All shortcomings of \ac{DFT} result from  the approximate nature of $E^{\rm XC}$. 
Most notable is the so-called self-interaction error.
In \ac{HF}, the interaction of an electron with itself is canceled  by the exact relation $J_{ii}-K_{ii}=0$.
This is, however, not true anymore if approximate \ac{XC} functionals are considered. This appears to be critical in many respects \cite{Kuemmel_RMP_2008} and especially for the core-levels of atoms (see Section \ref{sec:reponse} below).
%
\subsubsection{(Quasi)-one-electron methods for X-ray spectroscopy}
%
The \ac{KS-DFT} method has replaced HF in modern applications. Although  being also of one-electron nature, it describes independent quasi-particles, implicitly accounting for (dynamic) electron correlation.
Both methods can be used to generate a suitable basis of orbitals for subsequent  electron correlation treatment by single-reference methods (Figure~\ref{fig:overview}A), such as \ac{CI}, \ac{PT}, or \ac{CC} approaches as discussed below. 
The choice  of a proper orbital basis is more important for core-excited states than for valence-excited ones.
The reason is that removing an electron from the core region where the density of electrons is very high significantly influences electrons occupying other shells due substantial changes in the screening of the nuclear charge, which can be associated with an additional polarization potential \cite{Hedin_JPB_1969}.
Thus, most of the orbitals will strongly relax to become more tightly bound in energy and more confined in space.
Naturally, this type of orbital relaxation is very important for core ionization or the description of the subsequent \ac{NXES}; although there might be exception, see Reference~\citenum{Smolentsev_JACS_2009}.
This may require to perform an \ac{SCF} orbital optimization to account for the core-hole excited state to accurately predict X-ray spectra.
The simplest way to do so is to obtain relaxed orbitals where the core-excited atom is replaced by the next element from the periodic table. Within the so-called ${Z+1}$ \textit{or equivalent core approach} \cite{Jolly_JACS_1970} these orbitals are used for calculations of the parent molecule. Important applications include the interpretation of \acp{PES} and thermodynamic data \cite{Agren_CPL_1988}.

Although both \ac{HF} and \ac{DFT} techniques were described in the previous section as purely ground state methods, in principle, they can be applied to excited states as well.
In this case, the excited state is also represented as a single Slater/Kohn-Sham determinant analogously to the ground state; this is called $\Delta$\textit{SCF method} since the transition energy is be obtained as the difference between respective \ac{SCF} energies.
Its application to core-excited states has been first suggested in Reference \citenum{Bagus_PR_1965}. Because some of the features of
the $\Delta$SCF method will be important for the later discussion of \ac{MCSCF} calculations of core-excited states,  more details will be provided in the following.

A major issue for all methods which involve orbital optimization (cf. Figure~\ref{fig:overview}B,D)
is the so-called \textit{variational collapse}.
Let us assume that we have setup a wave function with a hole in some core orbital and an additional electron in a virtual orbital and naively perform a variational \ac{SCF} optimization (note that no spatial symmetry is assumed and the spin coincides with that of the ground state).
For a randomly chosen initial orbital guess, the virtual orbital hosting an excited electron will be transformed to the core one in a series of unitary transformations at every \ac{SCF} iteration. In other words,
without additional tricks the wave function is likely to  collapse to the ground state and core-excited states cannot be accessed.
The only reason why such calculations of deep core-hole excitations might converge to the desired local minimum solution is that usually these deep-lying core orbitals have a negligible overlap and couplings with other \acp{MO}.
For larger molecules and shallow holes this, however, might not work leading to oscillations, divergence, or to collapse to the ground state.

To circumvent this problem, several schemes have been suggested (Figure~\ref{fig:overview}B). In \textit{approach I}, one  searches consecutively for higher-lying variational solutions, e.g., under the constraint that each next solution should be orthogonal to all previously found, although this additional condition is not always rigorously applied, see below.~\cite{Evangelista_JPCA_2013} 
This might be a tedious procedure if  more than a few core-excited states are needed.
Alternatively, in \textit{approach II}, a one-electron orbital basis is  constructed  in such a way, that the differences of orbital energies give good estimates of the X-ray transition energies ($\Delta \varepsilon$ \textit{approach}).

Several different schemes along the lines of approach I have been suggested  to perform a constrained search. They differ in the construction of  occupied and virtual \acp{MO} involved in the excitation and in setting their occupation numbers, see Reference \citenum{Evangelista_JPCA_2013} and references therein.  
Exemplarily, one should mention the application of a multiple hole/particle algorithm within orthogonality constrained \ac{DFT} to the calculation of $K$-edge spectra \cite{Derricotte_PCCP_2015}.
Moreover, state-following algorithms can be applied to avoid the variational collapse without explicit setting the orthogonality constraint.
Here, one specifies the desired occupation number pattern for the initial guess and attempts to maintain it during \ac{SCF} iterations. 
This can be accomplished by applying different kinds of projectors to separate core orbitals from others at every iteration, see Reference \citenum{Hsu_JCP_1976} and references therein. Alternatively, one can utilize a maximum overlap criterion \cite{Gilbert_JPCA_2008, Besley_JCP_2009} to identify the core-hole orbital at each SCF step. 
In principle, one can also apply conventional \ac{SCF} subject to a special construction of the initial guess as discussed in References \citenum{Jensen_JCP_1987,NavesdeBrito_JCP_1991}.
Here, in the first step the singly-occupied core orbital is frozen and thus it cannot mix with other \acp{MO} via unitary transformations $\hat U(\hat{\kappa})$, i.e.\ one gets a constrained \ac{SCF} minimum.
Next, one uses this partially relaxed wave function to converge to the local energy minimum, corresponding to a fully relaxed core, making use of quadratically converging algorithms.

In approach II, one aims at obtaining orbitals partially relaxed due to presence of the core hole such that differences of orbital energies, $\varepsilon_i$, are a good approximation of transition energies $E_i-E_g =\bra{\Phi_i}\hat{H}\ket{\Phi_i}-\bra{\Phi_g}\hat{H}\ket{\Phi_g} \approx \varepsilon_i - \varepsilon_g$.
Several versions have been suggested, differing in the included fraction of the core hole and excited electron causing orbital relaxation, e.g., full or half of the core hole and full neglect or inclusion of the excited electron \cite{Chong_CPL_1995, Prendergast_PRL_2006}. 
One of the prominent representatives of approach II is the \textit{\ac{TP-DFT}} method \cite{Triguero_PRB_1998}   reviewed  in Reference~\citenum{Leetmaa_JESRP_2010}.
Here one chooses a half-occupied core orbital and neglects the effect of the excited electron.
Such a choice of a partial occupation can be viewed from a more general viewpoint of ensemble \ac{DFT} \cite{Theophilou_JPC_1979, Filatov_DMfES_2015}, where the number of electrons in the system varies from $N-1$ to $N$.
For the exact \ac{XC} potential, the energy between points with different integer particle numbers $E(N-1)$ and $E(N)$ changes linearly with fractional occupation $n$ and the derivative of the energy exactly amounts to the orbital energy $\partial E/\partial n_i = \varepsilon_i$ at any value of $n_i \in (N-1,N)$ \cite{Perdew_PRL_1982}.
This is however not the case for approximate \ac{XC} functionals as the energy behaves nonlinearly.~\cite{Mori-Sanchez_PRL_2008}
According to  Lagrange's mean value theorem the exact relation between the energy derivative and $\varepsilon_i$ is restored at some  point between $N-1$ and $N$ electron numbers.
In a sense, \ac{TP-DFT} is pretending to search for this optimal point approximating it by $N-1/2$.
More generally, fractional occupation can be considered as a ``fitting'' parameter to generate optimal orbitals for X-ray calculations.\cite{Williams_JCP_1975, Ehlert_JCC_2017} 

Note that in a number of works an even   more simplified protocol belonging to approach II, the so-called \textit{\ac{FOA}}, has been used. Here, only ground state SCF is carried out and intensities and energies are fully based on transition matrix elements and energy differences between ground state Kohn-Sham orbitals. 
\ac{FOA} is thus completely neglecting excited state orbital relaxation and in this respect  it is similar to 	Koopmans' theorem approach for \ac{HF} \cite{Koopmans_Phys_1934} or its analogues for \ac{DFT} \cite{Perdew_PRL_1982, Almbladh_PRB_1985,Chong_JCP_2002} as applied to X-ray \ac{PES} \cite{Hedin_JPB_1969}.
However, it gives reasonably good results for $K$-edge \ac{XAS} and \ac{NXES} of \ac{TM} compounds \cite{Lee_JACS_2010, Lassalle-Kaiser_IC_2013}.    

For approach I, the calculation of transition matrix elements is quite complicated in case of non-orthogonal wave functions and can lead to spurious results.
Within approach II they are usually estimated on a quite approximate basis as $\mathcal{X}(\Omega) \propto |\bra{\Phi_i}\hat{\mathbf{d}}\cdot \mathbf{e}_{\rm in} \ket{\Phi_g}|^2 \approx |\bra{\phi_i}\hat{\mathbf{d}}\cdot \mathbf{e}_{\rm in} \ket{\phi_g} \times S|^2$.
Here, $\ket{\Phi_g}$ and $\ket{\Phi_i}$ are SCF solutions for ground and $i$th core-excited states and $\ket{\phi_g}$ and $\ket{\phi_i}$ are orbitals
corresponding to the respective one-electron excitation.
Thus, the transition matrix element between $N$-electron states is replaced by a one-electron integral.
The factor $S$ is the \bt{$N-1$}-electron overlap which accounts for the relaxation of other orbitals not involved in the excitation.
It is usually assumed to be constant, but can be calculated in a more elaborate way to improve computed intensities \cite{Liang_PRL_2017}.

The main advantage of the $\Delta$SCF method is that one can use the well-developed machinery  of  ground state SCF theory for obtaining excited state properties without much of new development. One should keep in mind, however,  that in case of DFT-based techniques the accuracy of the underlying density functional is of vital importance. 
$\Delta$SCF-like methods have been widely applied to the $K$-edges of the second period elements from small molecules \cite{Triguero_JESRP_1999, Takahashi_JCP_2004, Besley_JCP_2009} to macromolecules \cite{Ferre_JCP_2002, Loos_IJQC_2007} and condensed phases \cite{Hetenyi_JCP_2004, Prendergast_PRL_2006, Leetmaa_JESRP_2010, Liang_PRL_2017}.
It has been even attempted to calculate energies of $L$-edge transitions of Si, P, S, Cl, and Cu containing systems \cite{Besley_JCP_2009}, although without proper treatment of relativistic effects like SOC.
Finally, $\Delta$SCF can also be used  to optimize electronic states with multiple holes residing on the core or valence \acp{MO} as relevant for Auger spectroscopy and shake-up \ac{XAS} transitions.~\cite{Agren_CPL_1975, Svensson_CPL_1976}
\subsection{Methods including correlation}
\subsubsection{Configuration interaction}
\label{sec:CI}
Apart from $\Delta$SCF-like approaches adapting the  \ac{HF} or \ac{KS-DFT} ground state to compute higher-lying states, there is a plethora of single-reference energy-domain excited state methods; cf. Figure~\ref{fig:overview}A,B.
The respective excited states  can be  represented in the general  form
\begin{equation}\label{eq:excited_general}
\ket{\Psi_{i}}=\mathcal{\hat E}_{i} \ket{\Psi_g} \, ,
\end{equation}
where $\ket{\Psi_g}$ is the ground state obtained with any suitable method and $\mathcal{\hat E}_{i}$ is an excitation operator.
In single-reference methods, $\ket{\Psi_g}$ is usually taken to be a single ground state Slater (or Kohn-Sham) determinant, $\ket{\Phi_g}$, and orbitals are not further optimized. We start the discussion of this class of methods with the \ac{CI} technique, being conceptually the simplest although not computationally most efficient representative.
The \ac{CI} framework facilitates understanding of other (even conceptually different) methods as it provides a very pictorial tool of wave function composition.

\ac{CI} is one of the most logical  ways to improve on the \ac{SCF} description \cite{Shavitt_CI_1977, Sherrill_AQC_1999}.
First it provides a systematic way for inclusion of electron correlation and second it can be naturally extended to the calculation of excited states.
Since ground state and excited Slater (or Kohn-Sham) determinants form a complete basis, 
a CI state $i$ can be written as
\begin{equation}\label{eq:CI_WF}
\ket{\Psi_{i}^{\rm CI}}=\sum_j C_{ij} \ket{\Phi_j(\{\phi_k^{\rm fixed}\})} \,.
\end{equation}
Here, the determinants $\ket{\Phi_j}$ are built from the fixed \ac{SCF} orbitals $\{\phi_k^{\rm fixed}\}$,  obtained prior to the \ac{CI} procedure, e.g., by ground state \ac{HF} or \ac{KS-DFT}. 
In practice, the coefficient matrix $\mathbf C$ is obtained 
by writing the electronic Hamiltonian in the basis of determinants which is iteratively diagonalized using the Davidson algorithm \cite{Davidson_JCP_1975}.

Using all possible determinants in the expansion Equation~\eqref{eq:CI_WF}  corresponds to the exact \ac{FCI} solution for a given basis set.
Its computational complexity grows exponentially with the number of basis functions and truncation schemes are usually applied. Starting point is the classification
 of the terms in Equation~\eqref{eq:CI_WF} as being unexcited reference, singly, doubly, triply, and so on excited electronic configurations:
\begin{equation}\label{eq:CI_level}
\ket{\Psi^{\rm CI}}=\underbrace{\ket{\Phi_g}}_\text{ref.} + \underbrace{\sum_{ia} C_{ia} \ket{\Phi_i^a}}_\text{singles} + \frac{1}{4}\underbrace{\sum_{ijab} C_{ijab} \ket{\Phi_{ij}^{ab}}}_\text{doubles}+...
\end{equation}
In this notation, indices $i,j,...$ and $a,b,...$ denote occupied orbitals from where an electron is excited and virtual orbitals to which electrons are excited in the determinant $\ket{\Phi_{ij\ldots}^{ab\ldots}}$, respectively.
To make the problem computationally tractable, one truncates the expansion Equation~\eqref{eq:CI_level} including, e.g., all determinants up to a given excitation class.
This gives rise to \ac{CIS}, \ac{CISD} methods, and so on.
Since electron correlation is at least a two-body effect, doubles are necessary to correct for it on top of a \ac{HF} reference.
Singles are important for description of charge density and thus for properties like transition dipole moments.
If only singles are included, it might provide a good tool for spectroscopy with near-\ac{HF} quality for the excited states.

From the viewpoint of quantum chemistry the core-excited electronic states are not any different from the valence-excited ones.
But, the  problem is that   one would need to calculate thousands or even millions of states to reach the highly lying levels relevant for X-ray spectroscopy. 
This problem stems from the fact that, if one is interested in the $n$th eigenvalue, the efficient Davidson algorithm \cite{Davidson_JCP_1975} standardly used for this purpose, diagonalizes at least an  $n \times n$ matrix and thus obtains non-interesting eigenvalues below $n$.
Since matrix diagonalization is the main computational bottleneck such a brute force approach wastes computational resources if this is possible at all.
Let us consider, for instance, two systems H$_2$O ($K$-edge) and [Fe(H$_2$O)$_6$]$^{2+}$  ($L$-edge).
If one takes a quite moderate triple-exponent TZP basis  and accounts for singly and doubly excited configurations which is the minimal reasonable choice to include electron correlation within the \ac{CI} approach, the total amount of valence states is 46\,872 and 136\,370\,360, respectively.
This means that to reach the first $1s$- or $2p$-core-excited state one would need to explicitly calculate a huge number of states, where in case of \ac{XAS} often only few low-lying ones are of relevance for the spectrum (due to the factor $f(E_g, T)$ in Equation~\eqref{eq:XAS}).

\begin{figure}[tbh]
	\centering
	\includegraphics[width=0.95\textwidth]{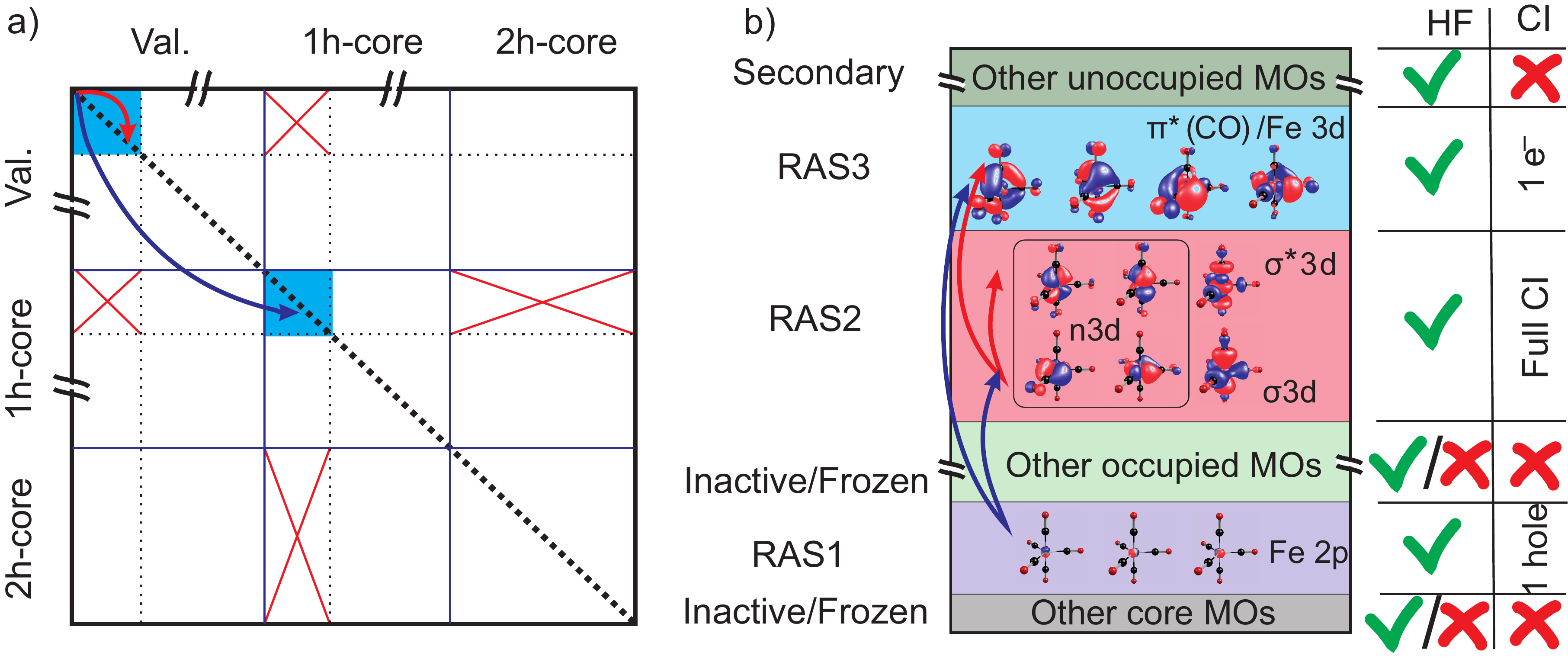}
	\caption{\label{fig:theory}
		 (a) Illustration of the projection of the full Hamiltonian to the subspace of limited valence- and core-excited configurations (blue blocks), neglecting the respective off-diagonal coupling blocks (red crosses).  %
		 (b) Principal \ac{MO} subspaces used in the \ac{RASSCF} method.  The typical \ac{AS} is exemplified for the Fe(CO)$_5$ complex. 
		 The table on the right demonstrates the level of orbital optimization (\ac{HF}) and \ac{CI} for each subspace. (Valence electronic excitations are depicted by red and core excitations by blue arrows, respectively.) %
            }
\end{figure}

There are several ways to circumvent this difficulty.
First of all, one can stay within the conventional Davidson scheme and project  onto a smaller subspace of trial functions excluding most (or even all apart from the ground state) valence states (Figure~\ref{fig:theory}).
In practice one   makes use of the concept of an \ac{AS} of \acp{MO}, where certain orbitals (i.e. the core orbitals for XAS calculations) are allowed to change their occupations, while for the others electronic excitations are forbidden (also called
\ac{CVS} \cite{Cederbaum_PRA_1980,Barth_PRA_1981, Schirmer_PRA_1990}).
Equivalently, the subspace of electronic configurations can be limited according to simple Koopmans' estimates of the state energy based on the orbital energies as realized within  \ac{REW-TDDFT}   \cite{Stener_CPL_2003, Ray_CEJ_2007}.

This ``reduced subspace'' concept is illustrated in Figure~\ref{fig:theory}a.
It shows a pictorial view of the CI-like Hamiltonian matrix  expressed in the basis of electronic configurations resulting from valence,  single core-hole (1h-core), and double core-hole (2h-core) excitations.
The overall dimensions of the Val--Val block are those thousands or millions mentioned before.
In fact, in \ac{XAS} one is usually interested in relatively small sub-block (marked with blue color) of the 1h-core--1h-core block including only few tens, hundreds or even thousands of states if only the discrete pre-edge region is of interest.
In case of \ac{RIXS}, one needs in addition a number of low-lying valence-excited states also denoted as a blue sub-block of the Val--Val block.
The 2h-core--2h-core block is usually completely neglected.
The same holds for the coupling Val--1h-core and 1h-core--2h-core blocks.
The motivation for neglecting off-diagonal couplings between two configurations from $v$ (valence) and $c$ (core) is provided by \ac{PT}. 
Here, the  error is estimated to be of the order of $V_{vc}/(E_c-E_v)$ in the wave function and $|V_{vc}|^2/(E_c-E_v)$ in energy, where $V_{vc}$ is the off-diagonal Hamiltonian matrix element and $E_c-E_v$ is the usually large difference between energies of these configurations standing on the matrix diagonal.  
In fact, the restricted subspace approach  is equivalent to neglecting specific type of two-electron integrals of the form $\braket{ab|cd}$, where three of the orbitals $a,b,c,d$ are valence and one is core orbital or vice versa and where orbitals in the bra are both core and in the ket  both valence \cite{Cederbaum_PRA_1980,Barth_PRA_1981, Schirmer_PRA_1990}.
Hence the diagonalization of the enormously large CI matrix in Figure~\ref{fig:theory}a can be replaced by a Davidson diagonalization of the small (blue) sub-blocks.
The uniform error introduced by \ac{CVS} was estimated to be of the order of 0.5-1.0\,eV for $K$-edges of the second period elements \cite{Barth_JPB_1985, Trofimov_JSC_2000}.

Another way to   access core-states is to modify the Davidson algorithm itself. For instance,
in Reference \citenum{Butscher_JCP_1976}, a modification of the standard algorithm \cite{Davidson_JCP_1975} employing the so-called ``root-homing'' technique has been suggested.
It utilizes the maximum overlap criterion (similar to $\Delta$SCF) to identify the desired state during each Davidson iteration step.
A more recent modification \cite{Liang_JCTC_2011}  builds on the idea of Reference~\citenum{Stratmann_JCP_1998} and
introduces an intermediate projection of the matrix to be diagonalized onto the subspace of the trial vectors having energies in a predefined window. After diagonalization the eigenvectors are projected back onto the full space.
The approximate energy of the trial vector is estimated as the orbital energy difference.
This approach is superior to the simple index fixing used in \ac{REW-TDDFT} \cite{Stener_CPL_2003} as it systematically includes contributions out of the initially predefined reduced space in a consistent way.
This gives a good approximation for  the high-lying states in the full space of trial vectors without explicit calculation of the lower-lying ones.
For the high density of states typical for X-ray spectra, a more advanced adaptive hybrid algorithm can be applied to improve the convergence \cite{Kasper_JCTC_2018a}.
Apart from \ac{CI} \cite{Butscher_JCP_1976, Buenker_CPL_1977}, so far such approaches were realized for \ac{LR-TDDFT} \cite{Liang_JCTC_2011, Lestrange_JCTC_2015, Kasper_JCTC_2018a} and \ac{EOM-CC} \cite{Peng_JCTC_2015} methods discussed below.

In the truncated \ac{CI} ansatz, Equations~\eqref{eq:CI_WF} and~\eqref{eq:CI_level}, the underlying one-electron orbitals determine accuracy and convergence of the  \ac{CI} expansion. That is why the choice of the orbital basis is of considerable importance for the overall efficiency.
It is known that virtual \acp{MO} from  ground state \ac{HF} calculations are too diffuse especially for large basis sets and thus represent a poor choice. 
The suggested remedies are to use improved virtual orbitals \cite{Potts_JCP_2001}, natural orbitals \cite{Thunemann_IJQC_1977}, or even \ac{KS-DFT} orbitals \cite{Hupp_CPL_2002}. 
In Reference \citenum{Ehlert_JCC_2017}, different types of the relaxed initial orbitals (restricted, unrestricted, and restricted open-shell \ac{HF} and \ac{KS-DFT}) have been considered and their performance for core-excitations within \ac{CIS} is deduced by comparison to experiment. 
It was found that a HF reference where orbitals are relaxed  to the core-hole is  most suitable.
Ground state frozen orbitals are not that good and produce errors of order of 10\,eV in core excitation energies.

\ac{CI} has been traditionally used for predicting X-ray spectra due to $K$-edge transitions in small molecules.
\cite{Butscher_JCP_1976, Buenker_CPL_1977, Barth_CP_1980, Cambi_CPL_1982}
For $L$-edge spectra of \ac{TM} complexes, the \textit{\ac{DFT-CI}} approach enjoys great popularity.
In this method, the standard \ac{CI} $\mathcal{\hat E}^{(i)} $ operator is applied to the ground state \ac{KS-DFT} determinant ($\ket{\Phi_{\rm KS}}$).~\cite{Marian_WIRCMS_2018}
The idea is to take the dynamical correlation effects at the level of \ac{DFT} into account and to incorporate stronger (static) correlation effects by \ac{CI}. As a result, the number of important determinants and thus the size of the Hamiltonian matrix can be substantially reduced.
This method largely relies on the better suitability of Kohn-Sham orbital energies as estimates of the electronic transition energies.
Since these orbital energy differences are dominating the diagonal of the \ac{CI} Hamiltonian matrix in the basis of determinants, it makes the major contribution to the final \ac{CI} energies \cite{Grimme_CPL_1996}.
However, since the \ac{CI} is built on top of quasi-particle orbitals which already implicitly include dynamic electron correlation a number of points have to be considered.  First, double counting of correlation effects should be excluded which can be done by energy-dependent exponential scaling of the \ac{CI} couplings between determinants \cite{Grimme_JCP_1999}.
Moreover, the two-electron integrals calculated on \ac{KS-DFT} \acp{MO} are overestimated and need to be scaled down even in case of \ac{CIS}, where no double counting occurs due to its ``uncorrelated'' character.~\cite{Grimme_CPL_1996, Roemelt_JCP_2013}
The need to parametrize this scaling makes the method semi-empirical in nature although the number of parameters is kept as small and universal/transferable as possible.
The parameters determined, e.g., by least-squares fitting to experiment, depend on many aspects: \ac{XC} functional, multiplicity and character of the target excited states, and so on.
Thus, the variants usually applied to valence \ac{TM} spectroscopy \cite{Escudero_JCP_2014, Marian_WIRCMS_2018} need to be adjusted to address core spectra, e.g., by  changes of the virtual orbital energies.~\cite{Grimme_CPL_1996}

The hybrid nature of  \ac{DFT-CI} provides another distinct advantage making use of the correlated origin of the Kohn-Sham \acp{MO}.
For \ac{TM} complexes with covalently bound ligands it reproduces the effect of bonding better than \ac{HF} which predicts bonds to be in general too ionic \cite{Neese_CCR_2007}. 
There are two   variants of \ac{DFT-CI} which were applied to X-ray spectra of \acp{TM}. In their approach, Ikeno et al. study crystalline metal oxide phases within a cluster approach.\cite{Ikeno_PRB_2005, Ikeno_U_2006, Kumagai_PRB_2008, Ikeno_JAP_2016, Ikeno_PRB_2011, Kumagai_JPCM_2009}  No corrections for double-counting or scaling of two-electron integrals is performed. 
As an \ac{XC} functional the simplest local density approximation is applied, thus containing no exact exchange.
Note that the \ac{CI} problem is solved only in small subspace of \acp{MO} containing few electrons (essentially $2p^{6-n}3d^{m+n}$, where $n=0,1$, and also $2p$ ligand orbitals to account for charge-transfer effects) and the other electrons and their interaction with active ones are taken in an averaged approximate way \cite{Watanabe_MSEB_1989}.
This makes this approach being strongly related to the \ac{LFM} semi-empirical method described below, although no fitting of parameters beyond usual \ac{DFT} level is performed.
Despite of the absence of fitting parameters and the approximations introduced, this method demonstrates a remarkable level of agreement with experimental \ac{XAS} and other types of spectra. 
A similar  \ac{CI} approach based on state-averaged relativistic Dirac-Hartree-Fock orbitals (no \ac{DFT})   has been applied to X-ray $4f$ and $2p$ \acp{PES} of UO$_2$ \cite{Ilton_SS_2008} and MnO \cite{Bagus_PRB_2006}, respectively.

The second \ac{DFT-CI} method specifically designed to calculate \ac{TM} $L$-edge spectra is \textit{\ac{DFT}-based \ac{ROCIS}}~\cite{Roemelt_JCP_2013}. The restricted open-shell \ac{SCF} calculation ensures the proper reference ground state configuration with unpaired electrons which is typical for \ac{TM} complexes. Being a \ac{DFT-CI} technique, it needs rescaling of two-electron integrals to   additionally reduce electron-electron interaction similar to \ac{LFM}.   
In practice, three parameters which are universal across the periodic table are introduced: two scaling Coulomb and exchange terms on the diagonal of the \ac{CI} matrix and one common scaling parameter for all off-diagonal elements.
With this it puts emphasis on the diagonal of the \ac{CI} matrix containing basically the excitation energy for a given orbital pair.
These scaling coefficients are optimized only for two density functionals: B3LYP and BHLYP in a specific def2-TZVP(-f) atomic basis. 
Thus, this method is semi-empirical making use of a training set containing experimental spectra.
From the viewpoint of \ac{CI} it accounts only for single electron excitations thus being intrinsically of \ac{SCF} quality for the excited electronic states and not including any additional electron correlation on top of \ac{KS-DFT} \acp{MO}.
Although being a relatively crude approximation, it allows to keep the configuration space very compact at the  price of neglecting   shake up excitations.
Therefore, it can significantly underestimate the density of states in energy regions where such satellite bands are occurring.
On the other hand, this approach is  computationally very efficient, especially in connection with a local correlation treatment \cite{Maganas_JPCA_2018}. It has been applied to  models including up to several tens of \ac{TM} atoms \cite{Kubas_JCTC_2018}.
This method is one-component (see Section \ref{sec:Relativism}) but scalar relativistic effects and \ac{SOC} are included what is prerequisite to reproduce $L$-edge X-ray spectra.
Overall, this approach provides fairly good agreement with \ac{XAS} experiments at least for highly-symmetric systems \cite{Roemelt_JCP_2013}.

Finally, we briefly mention \textit{\ac{LFM} theory} as another semi-empirical \ac{CI}-like method. It has played an important role in understanding the complex structure of X-ray spectra during the last three decades.
Within the \ac{MO} paradigm, the covalent bonding situation is usually described  as a one-electron effect resulting in mixing of orbitals of bound atoms.
\ac{LFM} follows more the valence bond theory route, where these effects are taken on the level of many-electron \ac{CI} between configurations with localized electrons.

\ac{LFM} theory has its origin in ligand field theory developed back in 1950s \cite{Griffith_QRCS_1957}, but there are several modifications to accommodate the finer spectroscopic features of \ac{TM} complexes.
It is  an atomic theory focussing on an isolated metal ion with, e.g., a $3d^n$ configuration (corresponding to an electronic state $\ket{3d^n}$).
The in general complicated chemical environment is taken into account as a perturbation via an effective electric field that disturbs the spherical symmetry of an atom \cite{Cowan_book_1981,Thole_PRB_1985}. 
Since the effect of ligands goes beyond effective electrostatic interaction especially in case of highly covalent metal-ligand chemical bond, this model needs to be significantly extended.
The major extension is to incorporate sophisticated bonding contributions in terms of charge-transfer between metal and ligand \cite{Okada_JESRP_1992}.
For this purpose in the framework of X-ray spectroscopy, one needs to consider not only $\ket{2p^63d^nL^0}$ and $\ket{2p^53d^{n+1}L^0}$ electronic configurations with ``neutral'' ligand ($L^0$) and either fully filled $2p$-shell ($2p^6$) or having one core hole ($2p^5$), but also states where an electron is transferred, e.g., from the ligand to the $3d$-shell -- $\ket{2p^63d^{n+1}L^{+}}$ and $\ket{2p^53d^{n+2}L^{+}}$.

If one extends the number of basis functions in which the LFM Hamiltonian  is written to the size of a few or even tens of thousands electronic configurations, upon diagonalization one obtains a complex structure of eigenvalues called multiplets.
These multiplets originate from the manifold of atomic levels, which are additionally split in energy due to the ligand field and charge-transfer effect.

Often, \ac{LFM} theory serves as a model for semi-empirical fitting of a number of parameters to reproduce experiments (for an  overview see Reference \citenum{deGroot_Book}).
The Coulomb and exchange interactions within core and valence manifold as well as between them are parametrized via Slater-Condon integrals precalculated at the \ac{HF} level and tabulated for each element. Additionally, their numerical values are often reduced to 80\% or even smaller, to roughly account for electron correlation.
The $3d$ valence  and the $2p$ core \ac{SOC} is also introduced semi-empirically as a parameter within $LS$ or $jj$ coupling limits (see below) depending on the \ac{TM} ion considered. 
Another parametrized input for the Hamiltonian is the ligand field. 
In case of high symmetry \ac{TM} complexes, e.g., octahedral or tetrahedral, only one effective parameter is used. 
However, the number of ligand-field parameters grows when symmetry is lowered.
Finally, the charge-transfer parameters are adjusted to fit experimental results. The overall number of adjustable parameters  can become quite large, thus complicating the fitting procedure.
Due to its semi-empirical nature this theory represents a valuable tool to postanalysis of experiments but has only limited predictive power.
The transition energies are usually calculated with an accuracy not better than 1.0 eV but in some cases an impressive agreement with experiment has been achieved.\cite{deGroot_Book}

To parametrize more complicated ligand fields the necessary information can be taken from ab initio calculations.
The tandem of the simplicity of \ac{LFM} theory with the versatility of first principles approaches gave a rebirth to the well-established concept.~\cite{Ikeno_PRB_2011}  
Moreover, band structure codes can be utilized for such a parametrization in case of solids \cite{Haverkort_PRB_2012,Hariki_PRB_2017}.
%
\subsubsection{\acl{MCSCF} method}
\label{sec:MCSCF}

Multi-configurational methods have been introduced to describe bond dissociation, distorted geometries, near-degenerate ground states, and electronically excited states. 
During decades of application, these methods have acquired a reputation of being the ``golden standard'' for this type of problems. \cite{Wahl_MCSCF_1977, McWeeny_CPR_1985, Roos_LectNotes_1992, Werner_ACP_1987, Shepard_ACP_1987, Roos_ACP_1987, Olsen_ACP_2007,Szalay_CR_2012, Lischka_CR_2018}
From the 1980's on, when the first efficient and rapidly converging algorithms for \ac{MCSCF} have been developed,  also  $K$-edge core-level spectra for diatomics and small polyatomic molecules have been calculated.~\cite{Jensen_JCP_1987, Schwarz_CP_1987, Bacskay_IJQC_1987}
During the last years investigations of $L$-edge X-ray spectra of \ac{TM} compounds emerged, where it has been shown to be able to cope with highly excited states in multiconfigurational situations \cite{Josefsson_JPCL_2012,Bokarev_PRL_2013,Pinjari_JCP_2014}.

The \ac{MCSCF} method is a hybrid of the \ac{HF} and \ac{CI} approaches.
It employs the CI multi-determinantal ansatz, Equation \eqref{eq:CI_level}, for the many-body state,
\begin{equation}\label{eq:RASSCF_ansatz}
\ket{\Psi^{\rm MCSCF}_{i}}=\sum_j C_{ij} \ket{\Phi_j (\{\phi_k^{(i)}\})} \enspace ,
\end{equation}
with the important difference that not only the CI coefficients,  $C_{ij}$, are  variationally minimized, but also the set of coefficients $c_{kl}^{(i)}$, within the \ac{LCAO} ansatz 
\begin{equation}\label{eq:MOs}
\phi_k^{(i)}(\mathbf{r} )=\sum_l c_{kl}^{(i)} \chi_l(\mathbf{r} ) \enspace .
\end{equation} 
The \ac{MCSCF} wave function can be parametrized as 
\begin{equation}\label{eq:MCSCF_function}
\ket{\Psi^{\rm MCSCF}_{i}}=\exp(\hat{\kappa}^{(i)})\left(\sum_j C_{ij} \ket{\Phi_j (\{\phi_k^{(i)}\})}\right),
\end{equation}
where operator $\exp(\hat{\kappa}^{(i)})$ describes an orthogonal transformation of a set of \acp{MO} $\{\phi_k^{(i)}\}$ which are specific to state $\ket{i}$ \cite{Roos_ACP_1987}.
The energy functional can be expressed as
\begin{equation}\label{eq:MCSCF_energy}
 E_{i}^{\rm MCSCF}(\mathbf{C},\pmb{\kappa})=\min_{\mathbf{C},\pmb{\kappa}}\frac{\sum_{jk} C_{ij} C_{ik}\bra{\Phi_j}\exp(-\hat\kappa^{(i)})\hat H \exp(\hat\kappa^{(i)})\ket{\Phi_k}}{\sum_k C_{ik}^2}\, .
\end{equation}
Upon joint optimization of \ac{CI} and \ac{MO} coefficients, the one-electron \ac{MO} basis   adapts to the description of electron correlation effects.  
This allows to quite reliably account for static correlation (multiconfigurational character) and (near-)degeneracies of the electronic states, which are typical for TM compounds.

However, this simultaneous treatment of orbital relaxation and electron correlation leads to a considerable increase in complexity of the actual simulations.
The number of CI configurations and MO coefficients quickly increases with the number of atoms and the number of basis functions per atom.
The straightforward application of Equation~\eqref{eq:RASSCF_ansatz}   considering all orbitals on equal footing is rather demanding if not impossible already for  medium-sized systems. 
There have been attempts to individually select the most important configurations, e.g., via perturbation theory. \cite{Engels_CC_2001, Bytautas_CP_2009} 
A more robust and popular solution is provided by the concept of the \ac{AS}. 
At the most general level it corresponds to the disjoint subsets of \acp{MO} for which different levels of \ac{CI} expansions are performed. 
Historically, the most popular variant is the \ac{CASSCF} approach~\cite{Roos_CP_1980}, cf. Figure \ref{fig:theory}b.
Here, the orbitals are classified in three groups: (i) those which stay doubly occupied are called  {inactive}, (ii) those which are unoccupied,  {virtual} or secondary, and   (iii) active ones, forming a subspace where all possible distributions of some number of active electrons are allowed, thus corresponding to a subspace \ac{FCI} calculation.
\bt{It is important to note  that inactive orbitals are treated at the \ac{HF} level and that is why the scaling with the physical size of the system is not prohibitive. The exponential growth of the computational demands can be thus mitigated or even avoided by treating only the local correlated subsystem within the \ac{AS}.
For fine tuning with respect to  important configurations, one can introduce additional subspaces restricting the number of electrons which can be excited from or to them.
This leads to the \ac{RASSCF} method \cite{Olsen_JCP_1988, Malmqvist_JPC_1990} if the number of active subspaces is equal to three or, in general, to the \ac{GASSCF} method \cite{Fleig_JCP_2001, Ma_JCP_2011} or similar approaches \cite{Ivanic_JCP_2003}.}

To sum up,  \ac{MCSCF} unites the benefits of the \ac{SCF} and \ac{CI} allowing not only for orbital relaxation upon core-hole formation but also for strong correlation effects.
Therefore, it optimally suits the treatment of \ac{TM} systems which are well-known for static correlations.
\bt{Moreover, as we have seen for the case of \ac{CI} for core-level calculations, the most pragmatic and robust approach to core-excited states employs  restricted subspaces of orbitals and excitations between this subspaces. 
This naturally leads to the idea of using \ac{CASSCF} or \ac{RASSCF} for X-ray spectroscopy.}
It should also suit to probe with X-rays the excited state dynamics involving bond cleavage or conical intersections, being a native habitat of \ac{MCSCF}-based methods. 

\bt{The choice of  included orbitals in the \ac{AS} fixes electron configurations considered within \ac{CI} treatment.
This has to be done in a balanced way as in general correlation effects do not naturally split into strong and weak ones but continuously fill the range of importance.
Thus, the selection of the \ac{AS} has been for a good while considered to be an art.}
As of now there is a consensus on how to select a more or less optimal \ac{AS} in general non-pathological cases \cite{Roos_ACP_1987, Roos_ACP_1996, Pierloot_COC_2001, Pierloot_MP_2003, Veryazov_IJQC_2011}. 
The \ac{AS} should comprise: (i) orbitals which are involved in the excitation to a particular state $i$, (ii)
all orbitals hosting unpaired electrons,
(iii) strongly correlating orbitals -- usually respective bonding and antibonding \ac{MO} pairs,
(iv) \acp{MO} which are responsible for near-degeneracy, e.g., $\sigma$ and $\sigma^\ast$ if the chemical bond is dissociating or degenerate $d$-orbitals for \acp{TM} or $f$-obritals for lanthanides and actinides,
(v) more specifically in case of $3d$-\acp{TM}, $3s$ and $3p$ orbitals might be included for the elements in the beginning of a row to include semicore correlation. For the later elements of a row, additional set of $4d$ orbitals is required to recover radial dynamic correlation of the electrons in the non-bonding $d$-orbitals already in the AS (double-shell effect). 
Note that some of the rules might be redundant since, for instance, $3d$ electrons are strongly correlating and at the same time respective orbitals can be singly occupied and degenerate.
Naturally, to perform a computationally feasible calculation, the number of active \acp{MO} as well as active electrons needs to be kept as small as possible.

On the one hand, core-excited states can be considered as quite difficult from the viewpoint of selection of \ac{AS} since the number of $d$-electrons changes in course of excitation/deexcitation or (auto)ionization.
On the other hand, in practice this does not represent a problem.
Exemplarily, let us consider, how to choose the AS for a core state \ac{RASSCF} calculation of the Fe(CO)$_5$  model system \cite{Suljoti_ACIE_2013}, see Figure~\ref{fig:theory}b.
Although Fe(CO)$_5$ has  a high symmetry, it is not obeying atomic selection rules.
However, X-ray excitation involves a very localized part of the electronic system, having predominantly atomic nature. Thus, we can still use approximate angular momentum ($\Delta l =\pm 1$) selection rules. 
Therefore, to study dipole allowed $L$-edge transitions one needs to include at least $2p$ and $3d$ orbitals into the AS as illustrated in Figure~\ref{fig:theory}b. 
The AS is further subdivided into RAS1, RAS2, and RAS3 subspaces. 
Usually, one is interested in singly core-excited states whereas doubly core-hole states are not relevant (Figure~\ref{fig:theory}a). 
To account for this and simplify the calculation, $2p$ orbitals are put in a separate subspace {RAS1}.
Further, only one electron is allowed to be excited from {RAS1} to other active orbitals, meaning that {RAS1} contains 6 (for valence states) or 5 (for core states) electrons distributed in the three $2p$ orbitals. 
The $3d$ electrons are usually highly correlated what especially applies to Fe(CO)$_5$ being a covalent complex \cite{Pierloot_MP_2003}.
Therefore, one would prefer to include them in a separate subspace, {RAS2}, and construct \ac{FCI}-type configurations.
Apart from the doubly-occupied non-bonding $n3d$ orbitals which reside almost exclusively on the Fe atom, there is a pair of bonding/antibonding $\sigma3d$ and $\sigma^\ast3d$ orbitals which are essential for correlation treatment.
Moreover, one of the two main transitions in \ac{XAS} is due to the $2p\rightarrow\sigma^\ast3d$ excitation and in \ac{RIXS} the $\sigma3d\rightarrow2p$ relaxation plays an important role.~\cite{Suljoti_ACIE_2013}
Additionally, one can include ligand orbitals, e.g., in the {RAS3} subspace and limit the number of electrons which are allowed to be excited to them from {RAS1} and {RAS2}, e.g., to one.
It is reasonable to include $\pi^\ast({\rm CO)/Fe}3d$ orbitals which have a mixed $\pi^\ast({\rm CO})$ and non-bonding $3d$ character. 
These orbitals are involved in the $2p\rightarrow\pi^\ast{\rm (CO)/Fe}3d$ transitions giving rise to an intense absorption feature \cite{Suljoti_ACIE_2013}.
These empty ligand-dominated orbitals can be included to describe effects of backdonation and explicit charge-transfer excitations as analyzed for the Fe(CO)$_5\,$ \cite{Suljoti_ACIE_2013,Wernet_Nat_2015} and [Fe(CN)$_6$]$^{3-/4-}\,$. \cite{Engel_JPCB_2014,Pinjari_JCP_2014}
Remarkably, the radial nodal structure of these \acp{MO} resembles that of the ``double-shell'' $4d$ orbitals.
Thus, being correlated, or in other words allowing for higher than single excitations to {RAS3}, they might help to describe radial $3d$ correlation. 
However, if one focuses on reproducing transition strengths somewhat sacrificing accuracy in energy, single excitations should suffice \cite{Suljoti_ACIE_2013}.
If one is interested in other charge-transfer effects, one might add additional orbitals to {RAS1} or {RAS3} spaces.
All other occupied orbitals are set inactive and stay always doubly occupied.

The complexity of the selection of the AS  for \ac{TM} compounds strongly varies with the covalence of the metal--ligand bond.
One can state that covalent complexes correspond to particularly strong static correlation, whereas ionic complexes can be essentially singly-configurational \cite{Pierloot_COC_2001, Pierloot_MP_2003}.
In case of ionic complexes [M(H$_2$O)$_6$]$^{n+}$, the AS can be kept even more compact including just three $2p$ and five $3d$ orbitals ($t_{2g}$ non-bonding and two $e_g$ anti-bonding orbitals in octahedral symmetry) \cite{Bokarev_PRL_2013,Atak_JPCB_2013,Grell_JCP_2015,Bokarev_JPCC_2015,Golnak_SR_2016}.
Other filled orbitals with $3d$ character, for instance, the  bonding $e_g$ counterpart, can be added to describe further shake-up and charge-transfer transitions as exemplified for, e.g. [FeCl$_6$]$^{3-}\,$ \cite{Pinjari_JCP_2014} and [Mn(H$_2$O)$_6$]$^{2+/3+}$ complexes \cite{Bokarev_JPCC_2015}.
Inclusion of $4s$ orbitals into the \ac{AS} accounts for allowed $2p\rightarrow4s$ excitation, having, however, lower intensity without notable influence on $L$-edge spectra \cite{Bokarev_PRL_2013}.
It might be still important for accurate reproduction of experimental spectra \cite{Chantzis_JCTC_2018}.
 
As a byproduct of having both filled and unfilled (or half-filled) valence \acp{MO} in the \ac{AS} one obtains a fraction of valence electronic states. 
In fact, states with valence $3d\rightarrow3d$ character can be used to calculate \ac{RIXS} and \ac{RPES}.
That is why, one should always take care that not only core but also bright valence states (in a sense of core$\rightarrow$valence radiative relaxation) are represented appropriately.
This condition is automatically fulfilled for the \ac{AS} shown in Figure~\ref{fig:theory}b. 
Finally, by choosing the \ac{AS} one can switch on and off various effects, e.g., in addition to $3d\rightarrow 2p$ back relaxation one can account for dipole-allowed $3s\rightarrow2p$ decay by simply adding $3s$ orbitals to the \ac{AS} \cite{Golnak_SR_2016}.

As a note in caution: Imposing constraints on orbital subspaces does not in general case solve the problem of variational collapse. 
The matrix $\pmb\kappa$ in Equations~\eqref{eq:MCSCF_function} and~\eqref{eq:MCSCF_energy} contains terms which are interchanging active \acp{MO} with inactive and virtual ones.
Effectively, those orbitals, which correspond to the largest coupling elements in the Hamiltonian, will  stay in or enter the AS.
Thus, there is always a probability that the calculation will not converge to the desired local minimum of the energy (with core orbitals in the \ac{AS}) but to the global one, where core \acp{MO} are rotated out of \ac{AS}.
As mentioned already for the case of single-reference SCF, if the initial orbital guess is reasonable then, due to small overlap of $2p$ orbitals with the valence ones and using at least a second order  algorithm, the initial orbitals will tend to stay active and the system will fall into the closest local minimum of energy.
In our experience, this procedure works quite well for high-spin states if the orbitals optimized for the lowest valence excited states are used as a starting guess.
Nevertheless, a systematic convergence to the desired local minimum cannot be ensured and often fails for medium- and low-spin electronic states of \ac{TM} complexes.
A possible solution might be to use a core-hole relaxed initial guess as suggested by \AA gren et al. \cite{Jensen_JCP_1987, NavesdeBrito_JCP_1991}.

Non-desired rotations can also be suppressed by restricting the elements of $\pmb{\kappa}$ mixing $2p$ orbitals with others to zero, i.e. to perform a constrained \ac{SCF} search.
This approach is justified by the findings in References \citenum{Hedin_JPB_1969, Firsht_MP_1976}, i.e.  that a hole in some deep-core shell does not influence much other orbitals in the same shell and thus relaxation of $2p$ orbitals can be neglected (see also References \citenum{Jensen_JCP_1987, NavesdeBrito_JCP_1991}).
In addition, one can utilize symmetry arguments in case of centrally symmetric systems. \cite{Pinjari_JCC_2016, Norell_PCCP_2018}.

Another ``brute force'' solution is to freeze all occupied orbitals in the inactive subspace, while still allowing for rotations between RAS$n$ subspaces and virtual orbitals.\cite{Bokarev_PRL_2013} This approach still includes some orbital relaxation due to possibility to ``borrow'' some orbital contributions from the virtual space and redistribute them among active \acp{MO}.
Although it sounds quite restrictive, in fact, it produces results which agree with experiments fairly well as illustrated in Section~\ref{sec:Applications}. This approach  allows to efficiently compute a few thousands of the electronic states which are needed to cover all relevant valence and core levels of the neutral and ionized systems as depicted in Figure~\ref{fig:orb_state_picture}b.

In fact, doing an \ac{MCSCF} calculation produces non-orthogonal electronic states since \acp{MO} for different states $i$ are not orthogonal with all the negative implications for the calculation of the transition properties. 
It is connected to notable numerical effort as the ($\mathbf{C},\mathbf{c}$) set needs to be determined for each electronic state separately, whose number for core-excited states can easily reach few thousands. 
To circumvent this and improve convergence properties, the state-averaging procedure is routinely employed \cite{Werner_JCP_1981}.
Within this approach the averaged ensemble density and energy over several electronic states are obtained via variational minimization of the functional 
\begin{equation}
E_{\rm av}=\sum_i w_i E_{i}^{\rm MCSCF} \enspace ,\ \ \ (w_1\geq w_2 \geq ... \geq w_N),
\end{equation}
with respect to a common set of orbital rotations.
According to the most general formulation of the respective variational theorem by Gross-Oliveira-Kohn~\cite{Gross_PRA_1988}, the weighting coefficients should not increase with the state number.
In practice, the weights $\{w_i\}$ are usually taken to be equal.
Thus, the index $i$ for the set of orbitals in Equation~\eqref{eq:MCSCF_function} can be omitted as they are the same for all states $\ket{\Psi^{\rm MCSCF}_{i}}$ in the ensemble.
Having a  common set of \ac{MO} coefficients, a more balanced description of several states is achieved and the difference in energies is resembled by the \ac{CI}.
Such a procedure allows to obtain a set of orthogonal non-interacting electronic states.
However, via averaging, the orbitals loose their optimality for the \ac{CI}-expansion.
Nevertheless, averaging allows to attain quite good agreement with experiment.
The state averaging for core states has been suggested by McWeeny \cite{McWeeny_MP_1974,Firsht_MP_1976} and used to interpret results of \ac{XPS} and Auger spectroscopies.

On the one hand, due to the large number of core-excited states which usually need to be accounted for, state averaging is a highly advantageous procedure as it reduces the computational effort.
On the other hand, if the optimal orbitals for different states are substantially different, this state-averaging procedure becomes not eligible.
One might expect  such a situation if a number of valence and core states are optimized simultaneously since the former require no relaxation upon the influence of the core hole and the latter do.
(A separate averaging of valence and core states has been also suggested \cite{Sergentu_JPCL_2018}.)
The unexpected practical solution to this problem is that even if one does democratic averaging over all the states obtained within a given \ac{AS}, the results agree well with experiments.
The democratic average can be viewed as accounting for non-integer core-hole occupation somewhat similar to the \ac{TP-DFT} approach.
Thus, the orbitals represent some compromise between needs of valence and core states \cite{Engel_JPCB_2014}.
In passing we note that this reminds on  the procedure where all the irrelevant lower-lying states are averaged with the interesting higher-lying ones to improve convergence and mitigate variational collapse \cite{Werner_JCP_1981}.

\subsubsection{\ac{RASPT2}}
The description of electron correlation within \ac{MCSCF} can be considered as unbalanced since it accounts for near-degeneracy effects (static correlation) but in addition also partially for the dynamic correlation within the \ac{AS}.
The \acp{MO} outside the \ac{AS} are treated at the \ac{HF} level with the respective moderate costs, see Figure~\ref{fig:theory}b.
A more balanced description requires a more complete account for the dynamic correlation contribution and can be achieved via  \ac{PT} applied on top of   \ac{MCSCF}.
Here,  one considers the correlation in an approximate way due to, for instance, single and double excitations between {inactive}, {active}, and {virtual} orbital subspaces in case of \ac{PT} up to second order.
The second order energy correction reads  
\begin{equation}\label{eq:PT2}
E^{\rm MR-PT}_i=-\sum_j \frac{|\bra{\Phi_j}{\hat V}\ket{\Psi_{i}^{\rm MCSCF}}|^2}{E_j - E_{i}^{\rm MCSCF}+\delta} \enspace ,
\end{equation}
where $\ket{\Psi_{i}^{\rm MCSCF}}$ is the \ac{MCSCF} state, Equation~\eqref{eq:MCSCF_function}, to be corrected by \ac{PT}, and configurations $\ket{\Phi_j}$ with respective energies $E_j$  correspond to the single and double excitations mentioned above, ${\hat V}$ is the difference between the true Hamiltonian and its zero-order approximation, and $\delta$ is the level shift. Among different flavors of \ac{MR}-\ac{PT} \cite{Hirao_CPL_1992a, Hirao_CPL_1992, Angeli_JCP_2001, Angeli_CPL_2001, Granovsky_JCP_2011} differing in the definition of the zero-order approximation, \ac{CASPT2} \cite{Andersson_JPC_1990, Andersson_JCP_1992} and its restricted variant \ac{RASPT2} \cite{Malmqvist_JPC_1990} have enjoyed popularity for excited states of \ac{TM} complexes in general \cite{Roos_QMESCwCA_1995,Roos_ACP_1996,Pierloot_MP_2003} as well as   for \ac{TM} $L$-edge X-ray spectra.
A concise introduction to this method listing main peculiarities and inherent problems can be found in Reference~\citenum{Pulay_IJQC_2011}; for a more detailed discussion see also References~\citenum{Roos_ACP_1996, Roos_QMESCwCA_1995}.

At this point we would like to highlight two issues:   (i) \ac{CASPT2} or \ac{RASPT2} cannot correct for the effect of a too small AS.
In fact, usually a larger \ac{AS} than the minimal one is required to get accurate PT results.
This  is reflected in the \ac{AS} selection rules listed above, for instance, the necessity to include ``double shell'' $d$-orbitals.
In principle, this effect can be mitigated by using the multi-state version of \ac{RASPT2} \cite{Finley_CPL_1998}.
Here, the effective Hamiltonian is constructed in the basis of single-state \ac{RASPT2} wave functions and subsequently diagonalized giving a more precise description.
 However, it might be impractical for core states due to their large number.

(ii) The energy expression in Equation~\eqref{eq:PT2} can be subject to singularities due to the vanishing denominators, giving rise to the so-called intruder state problem.
This is especially severe for core-excited states which can easily get degenerate with some $\ket{\Phi_j}$ configuration.
The natural solution  is to increase the AS if possible. Another solution is to apply a real-valued level shift $\delta$ (Equation~\eqref{eq:PT2}) \cite{Roos_CPL_1995, Roos_JMST_1996}, which artificially lifts the degeneracy with the unwanted weakly-interacting intruder states.
An even more elegant solution is the introduction of purely imaginary level shifts \cite{Forsberg_CPL_1997}, which is in fact utilized for the core-level computations. 
Nevertheless, choosing a universal level shift which is good for a large number of states at the same time might be problematic and one has to stay at the \ac{RASSCF} level.
Perhaps the best  solution is known as \ac{NEVPT2}\cite{Angeli_JCP_2001, Angeli_CPL_2001}. It explicitly includes two-electron terms in the zero-order Hamiltonian and is manifested to be intruder state-free.
However, it has not yet been widely applied to X-ray spectra calculations, except the work in Reference~\citenum{Chantzis_JCTC_2018}.

Finally, we note that the effect of \ac{RASPT2} depends on the covalence of the complex \cite{Pierloot_MP_2003}, with relatively small effect for aqueous metal ions \cite{Josefsson_JPCL_2012,Bokarev_JPCC_2015} but large impact for systems, like metal hexacyanides, where the affordable AS is not big enough to include all important ligand-dominated \acp{MO} \cite{Pinjari_JCC_2016}. 

Recently, the \ac{RASSCF}/\ac{RASPT2}-based multi-reference approach to X-ray spectroscopy of metal complexes was shown to be quite efficient in unraveling the nature of different transitions in static \ac{XAS} \cite{Bokarev_PRL_2013, Pinjari_JCP_2014, Golnak_SR_2016, Pinjari_JCC_2016, Sergentu_JPCL_2018, Kubin_JPCB_2018, Kubin_CS_2018, Chantzis_JCTC_2018}, \ac{RIXS} \cite{Josefsson_JPCL_2012, Suljoti_ACIE_2013, Atak_JPCB_2013, Engel_JPCB_2014, Bokarev_JPCC_2015, Preusse_SD_2016, Kunnus_SD_2016}, and core \ac{PES} \cite{Klooster_CP_2012, Grell_JCP_2015, Grell_JCP_2016, Golnak_SR_2016, Norell_PCCP_2018a} as well as dynamics studies \cite{Wernet_Nat_2015, Moguilevski_CPC_2017, Wang_PRL_2017, Jay_JPCL_2018}.
Selected examples will be presented in the Section \ref{sec:Applications}.

\subsection{Time-domain approaches}
A conceptually different approach is the direct solution of the \ac{TDSE}  (for a  review, see Reference~\citenum{Goings_WIRCMS_2018}) employing a  time-dependent Hamiltonian, $\hat H(t)=\hat H_{\rm mol} + {\hat V}_{\rm int}(t)$, with the field-free molecular Hamiltonian $\hat H_{\rm mol}$ and the molecule-external field  interaction ${\hat V}_{\rm int}(t)$.
The state $\ket{\Psi(t)}$ can be written in some time-independent basis, e.g., $\ket{\Psi(t)}=\sum_i a_i(t) \ket{\Phi_i}$, with the temporal evolution recast in the form of time-dependent coefficients.
This expansion can be done on the level of orbitals, leading to the time-dependence in \ac{MO} coefficients.
It results in the TD-\ac{SCF} method, e.g., \acl{RT} \acl{TDDFT} RT-TDDFT in case of the \ac{KS-DFT} ansatz for the wave function.
If one assumes  time-independent electronic configurations (e.g., Slater determinants) built on time-independent \acp{MO} allowing \ac{CI} coefficients evolving in time one gets the  \ac{TD-CI} method.
Uniting both approaches results in the \ac{TD-MCSCF} technique \cite{Sato_PRA_2015}, which in the most general form can be also viewed as an \ac{MCTDHF} method \cite{Meyer_WIRCMS_2012}.

The advantage of solving the  \ac{TDSE}  is that core-excited states can be accessed right away by choosing the proper frequency of the oscillating external field. Further, for pulsed excitation one can tune the width of the frequency window to be investigated.~\cite{Lopata_JCTC_2012} Besides focusing on the absorption of light, shaping the envelope of the pulse  opens a way to discuss the excitation of intricate superpositions of eigenstates and quantum control of the system's dynamics in real time.

Various real time methods have been applied for simulations of the core spectra and dynamics in the core-excited states such as  \ac{RT}-\acs{TDDFT} \cite{Lopata_JCTC_2012, Kadek_PCCP_2015, Kasper_JCTC_2018}, \ac{TD-CI} \cite{Mendive-Tapia_JCP_2013, Bauch_PRA_2014, Li_PRL_2015}, non-Dyson \ac{ADC} \cite{Kuleff_PRL_2011}, real-time time-dependent \ac{EOM}-\ac{CC} \cite{Nascimento_JPCL_2017}, and $\rho$-TD-RASCI.\cite{Wang_PRL_2017, Wang_MP_2017, Wang_PRA_2018}

\subsection{ Response function methods}
\label{sec:reponse}
While real time-domain propagation methods are emerging, perturbative response function approaches enjoyed considerable popularity over the last decades.  Since it falls outside the main scope of the review, we will briefly discuss  non-\ac{SCF} single-reference methods only (for an extensive review, see Reference \citenum{Norman_CR_2018}).
Response approaches are based on a perturbation  expansion of the time-dependent observables in   frequency domain~\cite{Kubo_JPSJ_1957},
$\bra{\Psi(t)}\hat O \ket{\Psi(t)}=\bra{\Psi_0}\hat O \ket{\Psi_0}+ \sum_{\omega_1} R^{(1)}(\omega_1)E(\omega_1)e^{-\rm{i}\omega_1t+\gamma t}+\ldots$, where an external monochromatic field $E(\omega_1,t)=E(\omega_1)e^{-\rm{i}\omega_1t}+{\rm c.c.}$ is assumed. In lowest order only the term linear in the field is retained and the limit $\gamma \rightarrow 0$ provides the basis for   \ac{LR} methods.
For instance, the absorption spectrum is obtained from the  Fourier-transformed first-order response or correlation function $R^{(1)}(\omega_1)$ where the operator $\hat O$ is the dipole moment operator. 
Moreover, by using correlation functions involving different operators
 one can generate the full realm of various propagator methods \cite{Oddershede_ACP_1987}. The main advantage of the so-called polarization propagator methods and closely related Green's function approaches \cite{Prigogine_1977} is that the spectrum is determined directly from poles and residues of the propagator.
This allows to avoid explicit calculation of the initial and final states and thus to minimize the error in differential correlation.

Among the propagator methods the \ac{ADC} scheme \cite{Schirmer_PRA_1982} and in particular its second-order variant \ac{ADC}(2) has been actively developed recently. As far as 
 core level calculations are concerned, including \ac{CVS}, it goes back to the  1980's \cite{Cederbaum_PRA_1980, Barth_PRA_1981} with early applications to X-ray spectra of small molecules  reported in References~\citenum{Barth_JPB_1985, Schirmer_CP_1988, Trofimov_JSC_2000}.
More complex recent applications have been reported 
 for \ac{XAS} \cite{Wenzel_JCC_2014, Wenzel_JCP_2015}, \ac{RIXS} \cite{Rehn_JCTC_2017}, \ac{PES} \cite{Pernpointner_JCP_2009, Kryzhevoi_JCP_2009, Neville_JCP_2016} and in combination with the Stieltjes method or a grid-based representation of the ionized electron to \ac{AES} and \ac{ICD} \cite{Gokhberg_JCP_2007, Kuleff_PRL_2007}.  

The most widely used response function approach is 
\ac{LR-TDDFT} whose versatility in predicting  $K$-edge spectra of   molecules containing main group elements is well documented, for a review see Reference \citenum{Besley_PCCP_2010}.
In the $K$-edge case, the relevant electronic states are essentially single-configurational and thus can be safely treated by a single-reference method.
Being related to the \ac{CIS} method \cite{Dreuw_CR_2005}, \ac{LR-TDDFT} needs to incorporate the same tricks as discussed for the \ac{CI} method.
These include configuration space or energy window restriction \cite{Stener_CPL_2003, Ray_CEJ_2007} or modification of the diagonalisation procedure itself leading to an energy specific solver.\cite{Liang_JCTC_2011, Lestrange_JCTC_2015, Kasper_JCTC_2018a}

\ac{LR-TDDFT} is prone to  problems due to the so-called self-interaction error. It 
correlates with the overlap between the occupied and unoccupied orbitals,  corresponding to an electronic transition \cite{Peach_JCP_2008}.
This overlap is very small for core-to-valence and especially for core-to-Rydberg transitions \cite{Besley_PCCP_2009} due to different radial sizes of very confined core and notably more delocalized virtual valence and Rydberg orbitals.
Consequently, the energies of core excitations are usually substantially underestimated by \ac{LR-TDDFT} with standard \ac{XC}-functionals \cite{Besley_JPCC_2007}.
Similar to charge-transfer valence-valence transitions \cite{Bokarev_JCP_2012}, energies of core transitions were found to be sensitive to the amount of exact exchange in the \ac{XC}-functional \cite{Besley_JPCC_2007, Lestrange_JCTC_2015}.
Here, exact exchange refers to orbital-dependent exchange integrals, analogous to the \ac{HF} $K_{kl}$ term, but calculated using Kohn-Sham orbitals.~\cite{Kuemmel_RMP_2008, Baer_ARPC_2010}
As a remedy, functionals with an adjusted amount of exact exchange have been suggested, such as BH$^{0.57}$LYP  for   carbon $K$-edge transitions \cite{Besley_JPCC_2007} and the CV-B3LYP for $K$-edges of second period elements, where core regions are described by BHHLYP (50\% \ac{HF} exchange) and valence with B3LYP (20\% \ac{HF} exchange).~\cite{Nakata_JCP_2006}
Another, more gentle way is to split the Coulomb operator into short-range and long-range parts depending on the interelectron distance and switch exact exchange between different values with a smooth function.
Most important for core spectroscopy is the short-range correction as introduced in References~\citenum{Song_JCP_2008, Besley_PCCP_2009}, leading to a remarkable improvement of $K$-edge transition energies. 
For other correction methods designed for prediction of X-ray spectra, see also References \citenum{Imamura_IJQC_2007, Tu_PRA_2007, Verma_JCP_2016}.

\ac{LR-TDDFT} has been successfully applied to main group element $K$-edges \cite{Besley_PCCP_2010, Liang_JCTC_2011, Lestrange_JCTC_2015, Baseggio_JCP_2017} demonstrating good accuracy and predictive power. Attempts have also been made to calculate $L$-edges of \acp{TM}.~\cite{Stener_CPL_2003, Campbell_JCP_2004, Ray_CEJ_2007, Lopata_JCTC_2012}
In particular, the \ac{REW-TDDFT} method has been used to demonstrate different kinds of non-linear X-ray spectroscopies such as stimulated \ac{RIXS} to probe the dynamics of electronic and nuclear wave packets in the excited states. \cite{Mukamel_ARPC_2013, Zhang_TCC_2015}

Another way to treat correlation on top of \ac{SCF} is provided by the \acf{CC} technique.
In contrast to the linear form of the  \ac{CI} expansion, it employs an exponential ansatz for the excitation operator such that the ground state wave function can be obtained, e.g., from the \ac{HF} solution as $\ket{\Psi_g}=\exp(\hat T)\ket{\Psi_{\rm HF}}$.
The operator $\hat T= \hat T_1 + \hat T_2 + ...$ generates singly, doubly, and so on excited configurations similar to the \ac{CI} method. 
One also does a hierarchical truncation according to excitation level as in the truncated \ac{CI} discussed above.
Its immense success for ground state studies \cite{Bartlett_RMP_2007} suggests an extension to the excited electronic states.

The \ac{CC} approach can be generalized to treat excited states within the \ac{EOM}~\cite{Bartlett_WIRCMS_2012} or \ac{LR} \cite{Koch_JCP_1990} formalisms, giving the same energies but different transition moments \cite{Koch_JCP_1994}. 
The advantage over conventional \ac{CI}   is that the excited state wave function is not a linear combination
of the determinants but has an exponential counterpart inherited from the reference state.
Essentially, it can be obtained from Equation (\ref{eq:excited_general}), where $\ket{\Psi_g}=\exp(\hat T)\ket{\Psi_{\rm HF}}$  and $\mathcal{E}_i$ has the same structure as in the  \ac{CI} case.
The operator $\exp(\hat T)$ effectively accounts for higher-order excitations due to its exponential form and $\mathcal E_i$ treats differential correlation with respect to the ground state.

Both \ac{EOM} and \ac{LR}-\ac{CC} have been applied  to $K$-edge core spectra recently.
For example, \ac{EOM}-\ac{CC} with singles and doubles excitations has been applied on top of the \ac{HF} reference with a core-hole obtained with maximum overlap method for emission spectroscopy of water \cite{Besley_CPL_2012}.
Alternatively, the energy specific algorithm, as discussed in Section \ref{sec:CI},  was used together with \ac{EOM}-\ac{CC} for a set of molecules containing main group atoms \cite{Peng_JCTC_2015}.
Different approaches have also been  used to get   core-ionized potential energy surfaces in case of the  ClF molecule \cite{Bazante_CPL_2017}.
In Reference~\citenum{Nooijen_JCP_1995}, a two-step approach is utilized, describing separately the effect of core-hole relaxation and an excited bound electron by means of electron-attachment variant of \ac{EOM}-\ac{CC}. 
A simpler reduction of the excitation space is introduced for the \ac{LR}-\ac{CC} method.\cite{Coriani_JCP_2015,Myhre_JCTC_2016}

The \ac{EOM-CC} and LR-CC methods fail if the ground state cannot be described by a single-reference \ac{CC} wave function.
In contrast to   \ac{CI}, the extension of the \ac{CC} idea to the multi-reference case is non-trivial due to the problem with definition of a proper vacuum state leading to a number of technical complications, for review see Reference \citenum{Lyakh_CR_2012}.
However, various flavors of multi-reference \ac{CC} have been applied to core-spectra of small main group molecules recently.~\cite{Brabec_JCP_2012, Dutta_JCTC_2014, Sen_JCP_2018}

\subsection{Relativistic effects}\label{sec:Relativism}
%

Relativistic effects are traditionally considered for the valence chemistry of  heavy elements.~\cite{Pitzer_ACR_1979,Pyykko_ACR_1979}
However, they also play an important role for core electrons as they experience the (almost) unscreened Coulomb potential of the  nucleus. For instance, these effects make $s$ and $p$ core electrons even more tightly bound to nuclei resulting in an additional stabilization of the $s$ and $p_{1/2}$ levels and destabilization of the $p_{3/2}$, $d$, and $f$ ones.~ \cite{Rose_JPB_1978} This adds different  energetic shifts, e.g., to $L_3$-edge ($2p_{3/2}\rightarrow3d$) and $L_2$-edge ($2p_{1/2}\rightarrow3d$)  transitions.
Further,  \ac{SOC} is determining the characteristic shape of $L_{2,3}$-edge spectra.

A rigorous treatment of relativistic effects is provided by the Dirac equation, for reviews see References~ \citenum{Wilson_1988, Schwerdtfeger_part1_2002, Hess_2003, Grant_2007, Dyall_2007, Reiher_2009}. 
Here,  the wave functions are four-component vectors, corresponding  to  positive (electronic) and negative (positronic) energy states. Four-component theory, equipped with a proper electron correlation treatment, is without doubt the ``golden standard'' of accuracy in relativistic calculations. But, the considerable computational costs
limit their applicability to atoms or small (highly-symmetric) molecules.
Examples of four-component core calculations include (i) DFT-CI based calculations of MO$_6^{n-}\,$ (M=Ti, Fe, Ni, etc)~\cite{Ikeno_PRB_2005, Ikeno_U_2006, Kumagai_PRB_2008, Ikeno_JAP_2016, Ikeno_PRB_2011} and UO$_8^{12-}\,$\cite{Ilton_SS_2008} clusters, (ii) LR-TDDFT calculations of $K$- and $L$-edges of H$_2$S \cite{Ekstrom_PRA_2006} and Si, Ge and S halides \cite{Fransson_PCCP_2016}, and (iii) ADC-based investigations of  XeF$_n$ $N$-edge spectra \cite{Pernpointner_JCP_2005} as well as of  photoionization of (HI)$_2$ and (LiI)$_2$~\cite{Pernpointner_JCP_2006} and of MnO. \cite{Bagus_PRB_2006}

Accounting for negative energy states contributes little to the physical and chemical processes most relevant for molecules containing no very heavy elements. 
This situation is reflected by the fact that in case of low and medium nuclear charge $Z$ there are solutions of the Dirac equation having dominant contributions from positive energy states.
Various  techniques have been designed  to decouple both components such that the small component is implicitly included in calculation of the large component. This can be done in principle within any predetermined level of accuracy yielding so-called exact two-component methods.\cite{Peng_TCA_2012} 
Examples include the \ac{ZORA} \cite{Chang_PS_1986} and \ac{IOTC} methods \cite{Barysz_JCP_2002},	 which are often used for core-level calculations.
The transition from four- to two-component theory can also be done via the Douglas-Kroll-Hess protocol \cite{Hess_PRA_1985, Hess_PRA_1986} which applies a sequence of  unitary transformations eliminating the coupling order-by-order and being exact in the limit of an infinite number of transformations. 
This yields a block-diagonal Hamiltonian  represented as a perturbational sum, where in addition to block diagonalization the spin-dependent and spin-free parts can be separated, thus  building a basis for distinguishing scalar and, e.g., \ac{SOC} relativistic effects.
In practice, however, one transforms only one-electron terms  as the transformation of two-electron integrals is very involved and is often sacrificed for the sake of efficiency. This might lead to inaccuracies in $L_3/L_2$ splittings as discussed in Reference~\citenum{Kasper_JCTC_2018}.

Such a block-diagonalization of the Dirac Hamiltonian with subsequent neglect of the negative energy block leads to two-component schemes. Here,  \ac{SOC} is usually taken into account at the stage of the variational \ac{SCF} orbital optimization leading to the so-called $jj$-coupling limit. 
The justification behind is that for high-$Z$ elements \ac{SOC} becomes comparable to the largest two-electron integrals and hence significantly influences the one-particle \ac{MO} basis.
Examples of 2-component calculations include 
(i) ZORA DFT-CI based calculations of oxides \cite{Kumagai_JPCM_2009}, 
(ii)  ZORA LR-TDDFT calculations  of $K$- and $L$-edges of Ti complexes~\cite{Casarin_JPCA_2007,Fronzoni_CPL_2005}, of chemical shifts in $K_{\alpha1,2}$ emission lines of YbF$_n$ \cite{Shakhova_OS_2018} and  Nb oxide \cite{Lomachuk_OS_2018}, XAS of gas phase 
\ac{TM} oxochlorides \cite{Fronzoni_JPCA_2009},  short-chain oligothiophenes XAS and XPS 
~\cite{Baseggio_JCP_2017}, 
(iii) real-time  DFT calculations of XAS for \ac{TM} chlorides and oxochlorides 
\cite{Kasper_JCTC_2018}, 
(iv) RASSCF IOTC calculations of XPS of noble gases.~\cite{Barysz_MP_2014}
 
For compounds containing low and medium $Z$ elements the most pragmatic approach is to  neglect the spin-dependent part of the Hamiltonian in the two-component scheme, resulting in a one-component approach.
It corresponds to the usual solution of the Schr\"odinger equation with the difference that scalar relativistic effects are taken into account. The latter have been shown to be crucial for predicting the correct energetics of  X-ray spectra.~\cite{Ekstrom_PRA_2006, Roemelt_JCP_2013, Verma_JCTC_2016} Typically, scalar relativistic effects are treated  at the second-order Douglas-Kroll-Hess level.

For $L$-edge spectra the magnitude of \ac{SOC} is larger than typical energy level spacings and thus it must be taken into account. Within the one-component protocol \ac{SOC} is included as a perturbation of the spin-free problem in the framework of the so-called quasi-degenerate perturbation theory.\cite{Marian_RCC_2001}
Here, the \ac{SOC} Hamiltonian, $\hat{H}_{\rm SOC}$, is represented in the basis of spin-free spatial wave functions multiplied by an appropriate spin part.
The  \ac{SOC}-coupled eigenenergies and eigenstates are obtained via matrix diagonalization and have the form of linear combination of spin-free states, $\ket{\Psi^{{\rm SF},(S,M_S)}_{n}}$, with complex coefficients
 \begin{equation}\label{eq:SOC_states}
  \ket{\Psi^{\rm SOC}_{a}}=\sum_{n}\xi_{an}^{(S,M_S)} \ket{\Psi^{{\rm SF},(S,M_S)}_{n}} \, .
 \end{equation}
 
 In practice for core-level calculations, it is advisable to use as many states for the zero-order state basis set $\{\ket{\Psi^{{\rm SF},(S,M_S)}_{n}}\}$ as are available for a given \ac{AS}.\cite{Pinjari_JCC_2016} Note that since in this approach not the total momenta of the individual electrons are coupled as in the $jj$-limit, but rather the total angular momenta $L$ and spins $S$ of many-body electronic states, it is referred to as $LS$-coupling limit. 

The expectation values and transition matrix elements are then calculated for the  \ac{SOC}-coupled eigenstates 
$\{ \ket{\Psi^{\rm SOC}_{a}}\}$.
Therefore, formally spin-forbidden transitions may gain notable intensity in X-ray spectra which is   borrowed from the allowed ones via \ac{SOC}-mixing \cite{Bokarev_PRL_2013, Bokarev_JPCC_2015, Golnak_SR_2016}. 
In passing we note that this mixed nature of \ac{SOC} eigenstates is also reflected in the structure of the \acp{DO}, which  become complex orbitals having spin-up and spin-down components due to mixing of different multiplicities.~\cite{Grell_JCP_2015}

For such an a posteriori treatment of \ac{SOC} the operator $\hat{H}_{\rm SOC}$ in the Breit-Pauli form \cite{Bethe_1977} is commonly used.
It can be represented in terms of one-, $\hat{h}_{\rm SOC}(i)$, and two-electron, $\hat{g}_{\rm SOC}(i,j)$,  operators
As usual, multi-center two-electron terms require the most effort in the calculation of $\hat{H}_{\rm SOC}$ matrix elements in the basis of spin-free electronic states.
However,   they cannot be neglected (especially for light elements) without notable decrease in accuracy. 

Let us consider a matrix element of $\hat{H}_{\rm SOC}$ between two Slater determinants differing in the occupation of the $i$th and $j$th orbitals, whereas occupations of other orbitals, $\{n_k\}$, are the same, i.e. 
\begin{equation}
 H_{ij}=\bra{i}\hat{h}_{\rm SOC}(1)\ket{j}+\frac{1}{2}\sum_k n_k [\bra{ik}\hat{g}_{\rm SOC}(1,2) \ket{jk}-\bra{ik}\hat{g}_{\rm SOC}(1,2) \ket{kj}-\bra{ki}\hat{g}_{\rm SOC}(1,2) \ket{jk}].
\end{equation}

An efficient  mean-field type approximation is to fix the occupation numbers 
$\{n_k\}$ to, e.g., ground state \ac{HF} values \cite{Hess_CPL_1996}.
Further, a speedup is achieved when the molecular mean-field is approximated by the superposition of the mean-fields of the constituent atoms.
Accounting for the locality of \ac{SOC}, decaying much faster with distance than the Coulomb interaction, and the fact that the largest couplings are expected for core orbitals which are close to the respective nucleus, one can stay with one-center two-electron terms  only. This completely separates molecular \ac{SOC} into additive atomic contributions.
The $\{n_k\}$ are standardly calculated by the \ac{HF} method for neutral atoms.
This approximation is called  \ac{AMFI} \cite{Schimmelpfennig1996} and appeared to be quite accurate \cite{Marian_RCC_2001}.
Numerous applications in the field of X-ray spectroscopy confirmed that this technique provides remarkable accuracy not only for valence but also for core-electrons as is evidenced by the prediction of $L_2/L_3$ energy separations and general shape of XAS, XPS, and RIXS. 
It is the basis of the \ac{SOC} treatment in the ROCIS-DFT method \cite{Roemelt_JCP_2013} and of all multi-reference calculations discussed in Section \ref{sec:Applications}.

Finally, we comment on the atomic basis sets which can be employed for core-level calculations.
The usual way of implicit inclusion of relativistic effects via core-potentials \cite{Dolg_CR_2012} is naturally not suitable for this purpose.
In this approach the core electrons are removed from the respective atoms, essentially simplifying the nodal structure of the \acp{MO}, and their effect is accounted by an effective pseudopotential and only transitions between valence levels can be rigorously calculated. However, there have been reports of two-component calculations using pseudopotentials with an a posteriori restoration of the core nodal structure.~\cite{Shakhova_OS_2018, Lomachuk_OS_2018}  However, in general utilization of  all-electron basis sets is prerequisite for explicit calculations of X-ray spectra.
Moreover, basis sets need to contain very tight functions to describe the core region; to increase the accuracy one might even de-contract otherwise contracted core functions.~\cite{Josefsson_JPCL_2012}
For this purpose, basis sets specifically designed for correlation and at the same time relativistic treatment from the families cc-pwCVXZ-DK \cite{Balabanov_JCP_2005} and ANO-RCC \cite{Roos_JPCA_2005} are usually utilized.
For calculations of core spectra of compounds containing atoms from the first three periods the IGLO bases have also been developed.~\cite{Kutzelnigg_DaSC_1990} 
%
\section{APPLICATIONS}\label{sec:Applications}

\subsection{Unraveling  the  electronic  structure  at  the  metal-solvent interface}\label{sec:el_struct}
\begin{figure}[tbh]
	\centering
	\includegraphics[width=0.99\textwidth]{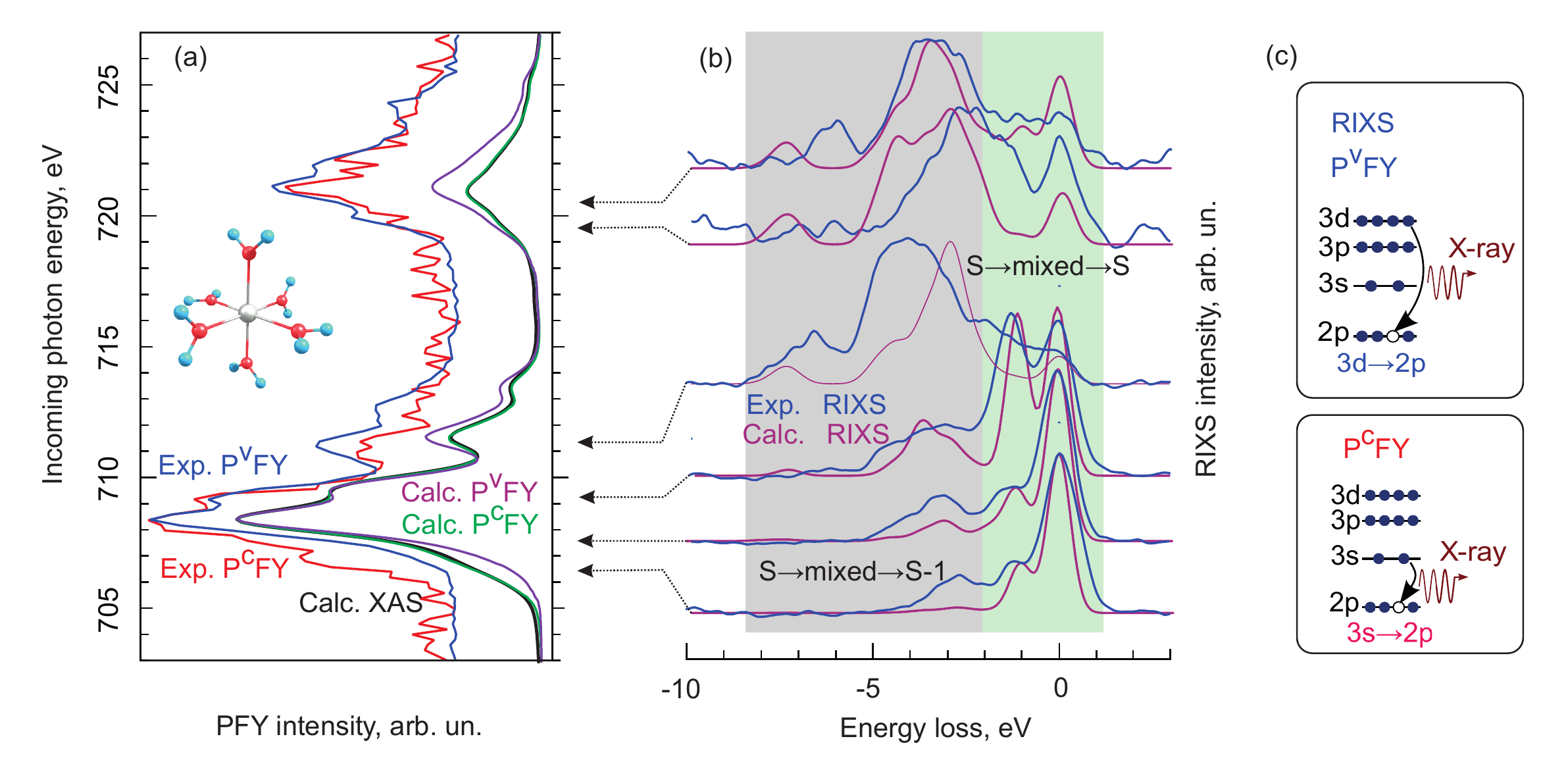}
	\caption{\label{fig:XAS_RIXS}
		 Experimental and calculated (\ac{RASSCF}/\ac{RASPT2}) spectra of the aqueous Fe$^{2+}$ ion (an [Fe(H$_2$O)$_6$]$^{2+}$ cluster is used in simulations, see inset). (a) Absorption spectra obtained in different modes \cite{Golnak_SR_2016}: true \ac{XAS} (calculated -- black), valence \ac{PFY} (Equation~\eqref{eq:PFY}) with $\omega \in [660,720]$\,eV denoted as P$^{\rm V}$FY (experiment -- blue, calculation -- magenta) and core \ac{PFY} $\omega \in [605,640]$\,eV denoted as P$^{\rm C}$FY (experiment -- red; calculation -- green). (b) Respective \ac{RIXS} spectra \cite{Atak_JPCB_2013} in energy loss ($\omega-\Omega$ in Equation~\eqref{eq:RIXS}) representation. The green and gray rectangles denote the energy range of the spin-allowed and  formally spin-forbidden transitions, respectively. (c) Orbital scheme showing different radiative relaxation channels. (adapted from References~\citenum{Golnak_SR_2016, Atak_JPCB_2013})}
\end{figure}

%
Solvated transition metal ions have been shown to provide a model system for metal-ligand interactions and in particular for the electronic structure at the metal-solvent interface
\cite{Bokarev_PRL_2013, Atak_JPCB_2013,Golnak_SR_2016}.
In Figure~\ref{fig:XAS_RIXS} XAS and RIXS spectra for the prototypical ferrous [Fe(H$_2$O)$_6$]$^{2+}$ complex are shown. This weak-field high-spin (quintet) complex has $d^6$ configuration and a Jahn-Teller distorted tetragonal symmetry.  The $d_{z^2}$ and $d_{x^2-y^2}$ orbitals of the iron ion form bonding $\sigma 3d$ and antibonding $\sigma 3d^*$ orbitals of $e_g$ symmetry with the $a_1$ orbitals of the water molecules. 
The other $3d$ orbitals of $t_{2g}$ symmetry form $\pi$-type  \acp{MO} with the water $1b_1$ and $1b_2$ orbitals, where the mixing is notably weaker as compared to the $\sigma$ orbitals.
\bt{That is why they can be denoted as non-bonding $n3d$ \acp{MO}. Note that symmetry labels assume octahedral symmetry for simplicity.}

Calculated and measured  absorption spectra are in rather good agreement as shown in Figure~\ref{fig:XAS_RIXS}a \cite{Atak_JPCB_2013, Golnak_SR_2016}. Focusing on the $L_3$ edge, analysis of the core-excitations reveals that almost all intense transitions are of $2p \rightarrow \sigma 3d^*$ type, but with substantial admixture of $2p \rightarrow n3d$ character. Notice that this mixing of $\sigma 3d(e_g)$ and $n3d(t_{2g})$ orbitals strongly depends on the considered ion. For Fe$^{3+}$, for instance, it has been found that the low-energy part of the $L_3$ edge is dominated by $t_{2g}$-type transitions \cite{Bokarev_PRL_2013}.  The multiplicity of the core-excited states strongly depends on energy \cite{Wang_PRL_2017, Wang_MP_2017, Wang_PRA_2018}. 
For energies up to about the $L_3$ maximum spin-conserving quintet-quintet transition dominate. The region up to 715~ eV is characterized by strongly mixed quintet-triplet transitions. The $L_2$-edge is comprised essentially of spin-forbidden quintet-triplet transitions, which borrow intensity by mixing with spin-allowed ones.

\begin{figure}[tbh]
	\centering
	\includegraphics[width=0.55\textwidth]{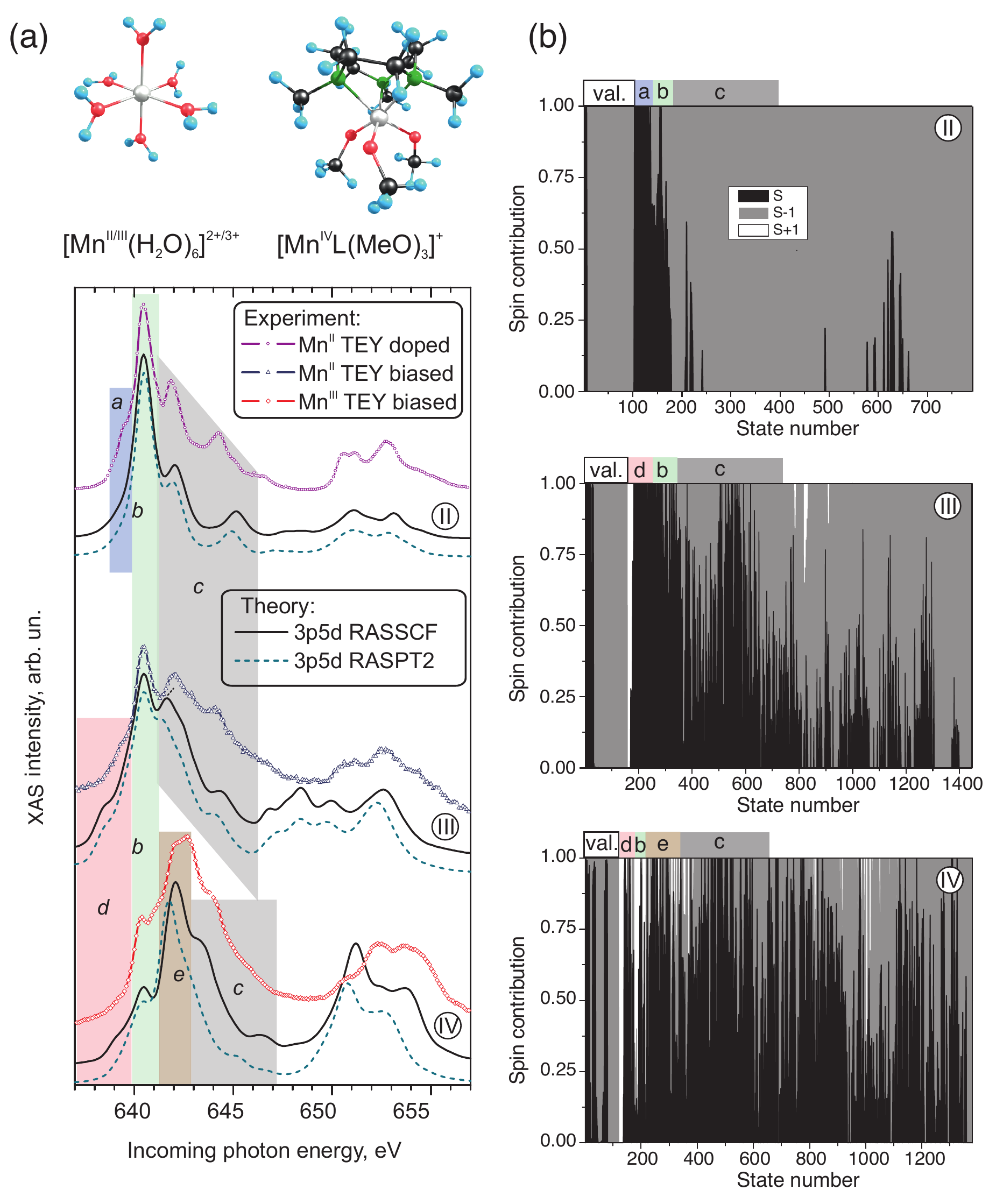}
	\caption{\label{fig:mn}
\rt{(a) Experimental and calculated (\ac{RASSCF}/\ac{RASPT2}) \ac{XAS} spectra for Mn species in different oxidation states.  (b) Contributions of states  with $\Delta S=0,\pm 1$ to the spin-orbit coupled wave functions for the different species of panel (a). The labels (val,a-e) correspond to the bands marked in panel (a) where the widths of the shaded areas comprise the state numbers of panel (b)  (adapted from Reference \citenum{Bokarev_JPCC_2015}).}} \end{figure}

\ac{RIXS} spectra of [Fe(H$_2$O)$_6$]$^{2+}$ shown in Figure~\ref{fig:XAS_RIXS}b can be used to characterize the electronic structure in more detail.~\cite{Atak_JPCB_2013} For excitation across the $L_3$-edge   up to its maximum,  most prominent is the elastic Rayleigh feature at the energy loss of 0\,eV. Inelastic peaks at about -1.2\,eV energy loss are due to emission from the doubly occupied $t_{2g}$ orbital and the respective shake-off satellites.
With increasing excitation energy the intensity of inelastic peaks also increases as compared to the elastic ones.  This evidences the fact that below -1.2\,eV valence excited states are of triplet character. 
Essentially, the excitation occurs from the ground quintet state with spin $S=2$ to a spin-mixed core-excited state (see also Section~\ref{sec:mn} and Figure~\ref{fig:mn}b).
The features with energy losses from 0 to -1.2\,eV correspond to $S\rightarrow {\rm mixed} \rightarrow S$ transitions.
Due to the mixing in the intermediate state, the new formally spin-forbidden radiative channels $S\rightarrow {\rm mixed} \rightarrow S-1$ are opened.
Since with increasing energy core-excitations are characterized by an increasing quintet-triplet  mixing, the elastic peak decreases and spin-forbidden features grow in intensity. 
\rt{Different spin states of iron complexes have been also studied in Reference~\citenum{Hahn_IC_2018}.}

The spectra shown in  Figure~\ref{fig:XAS_RIXS}b are essentially due to local metal-centered transitions. \ac{RIXS} features due to metal to ligand charge transfer are showing up around -15\,eV energy loss, but have small intensity. Hence, the mixing between the iron and water orbitals at the $L$-edge is of minor importance, \bt{evidencing the predominant ionic character of the metal-ligand bond}. Respective results obtained for probing the oxygen $K$-edge, yield rather similar spectra for [Fe(H$_2$O)$_6$]$^{2+}$  and for pure water, thus supporting the weak mixing (ionic) picture.~\cite{Atak_JPCB_2013}

Recording a true \ac{XAS} in transmission mode is a difficult task for optically-thick condensed phase samples, such as solutions, even if microjets are utilized.
Therefore, experimentally one usually obtains it in the \ac{PFY} detection mode, Equation~\eqref{eq:PFY} \cite{Sham_SRiMS_2018}.
The natural question arising in this respect is how large are the distortions of the \ac{PFY} relative to the true \ac{XAS} and how they can be  reduced.~\cite{deGroot_SSC_1994,	Kurian_JPCM_2012}
Figure~\ref{fig:XAS_RIXS}a shows the two experimental \ac{PFY} spectra resulting from $3d\rightarrow 2p$ (P$^{\rm V}$FY) and $3s \rightarrow 2p$ (P$^{\rm C}$FY) dipole-allowed radiative relaxation of the initially created core hole, \cite{Golnak_SR_2016} see panel (c).
The comparison of the calculated \ac{XAS} and \ac{PFY} spectra stemming from these two channels evidences that the core variant P$\rm ^C$FY should be a more reliable probe of \ac{XAS} in contrast to P$\rm ^V$FY.
The reason for the distortions of the latter has been identified as being    the opening of new radiative relaxation channels via strong \ac{SOC} in the core-excited state \cite{Golnak_SR_2016}, cf. discussion in previous paragraph for \ac{RIXS}.

The \ac{RIXS} spectra record photon-in/photon-out events and thus are determined by dipole selection rules as is evident from the Kramers-Heisenberg Equation~\eqref{eq:RIXS}. 
In Reference~\citenum{Golnak_SR_2016}, it has been demonstrated for [Fe(H$_2$O)$_6$]$^{2+}$ that a back-to-back analysis of \ac{RIXS} and photon-in/electron-out \ac{RPES}, which is free from such selection rules, provides complementary information on the electronic states. In particular, comparing intensities, one can obtain information on the competition between radiative and non-radiative decay channels of the core-excitation.
%
\subsection{ A new look at metal-ligand bonding}

\ac{XAS} and \ac{RIXS} spectroscopy applied to metal-ligand complexes can provide an atom-specific, chemical state selective, crystal field symmetry and orbital symmetry resolved description of the electronic structure \cite{Suljoti_ACIE_2013,Engel_JPCB_2014,Kunnus_JPCB_2016, Norell_PCCP_2018, Jay_JPCL_2018}. 
In References~\citenum{Suljoti_ACIE_2013,Engel_JPCB_2014,bokarev17_02004}, it has been shown that this allows for scrutinizing traditional chemical concepts of metal ligand bonding. Within the valence bond structure theory, metal-ligand bonding is described by the so-called $\sigma$-donor/$\pi$-acceptor mechanism. For example, the formation of the covalent Fe--CO bonds in Fe(CO)$_5$ is accompanied by charge donation from the $5\sigma$ orbital  of CO to the $\sigma3d$ orbital of the Fe and back donation to the $2\pi^*$ orbital of CO. In terms of the \acp{MO} of the metal-ligand complex, the donation/back donation is expressed by the mixing of the respective Fe and CO orbitals. The extent of this mixing can be addressed by identification of charge-transfer transitions in \ac{RIXS} spectra.

\begin{figure}[tbh]
	\centering
	\includegraphics[width=0.9\textwidth]{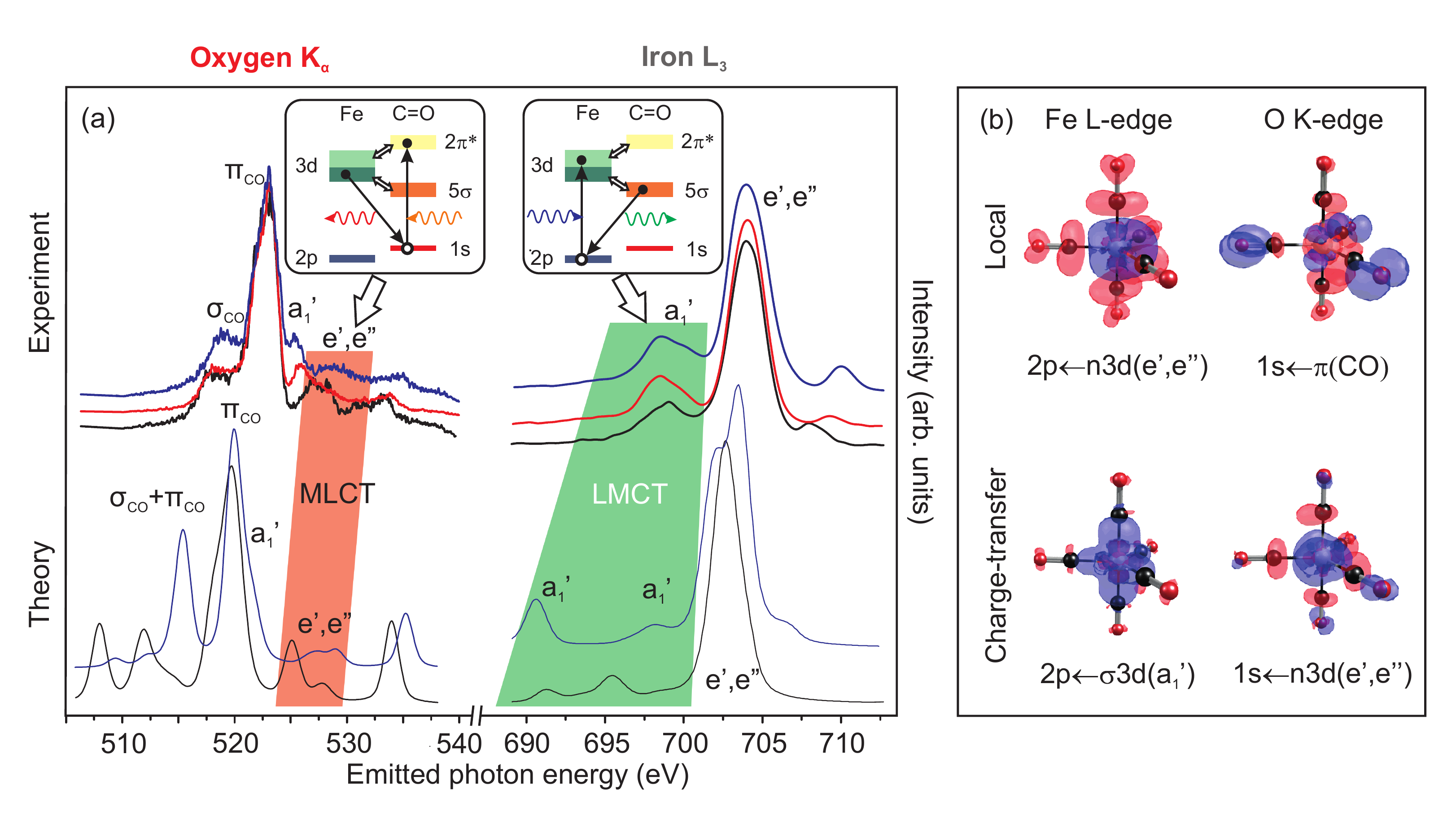}
	\caption{\label{fig:chemical_bonds}
		 (a) Experimental and theoretical \ac{RIXS} spectra of the Fe(CO)$_5$ complex at iron $L$-edge and oxygen $K$-edge. The colored areas show the range of metal-to-ligand (MLCT) and ligand-to-metal (LMCT) charge-transfer transitions which carry information about the strength of covalent bonding as explained in the respective insets.  (b) Transition density difference plots which show the localization of the excited electron (red) and the electron which is refilling the core hole (blue) within the complex. (adapted from Reference~\citenum{Suljoti_ACIE_2013})}
\end{figure}

For illustration, let us consider the \ac{RIXS} amplitude for the Fe $L$-edge in the simplified picture considering only \acp{MO}. The dipole operator can be written as
$\hat{d}=d_{ 2p \rightarrow 3d}  \ket{ 3d} \bra{ 2p}
+{\rm h.c.}$ Suppose that the $3d$ character of the absorbing state is given by the \ac{MO} coefficient $C^{\rm (abs)}_{ 3d}$ of a virtual orbital, to which the core electron is excited, i.e.
$\ket{\Psi^{\rm (abs)}}=\hat{d} \ket{2p}=d_{ 2p \rightarrow 3d} C^{\rm (abs)}_{ 3d} \ket{3d}$. 
Thus, core excitation projects out all \ac{MO} contributions apart from the $3d$ due to locality and relatively strict dipole selection rules.
The $3d$ character of the respective state for emission shall be given by the coefficient $C^{\rm (em)}_{ 3d}$ of the occupied orbital from which the core hole is refilled, i.e. $\ket{\Psi^{\rm (em)}}=\hat{d} \ket{ 3d}=d_{3d \rightarrow 2p} C^{\rm (em)}_{3d} \ket{ 2p}$. Hence, the \ac{RIXS} amplitude in Equation~\eqref{eq:RIXS} becomes
$\mathcal{R} \propto |d_{ 2p \rightarrow 3d}|^{4}  |C^{\rm (abs)}_{ 3d}|^2 |C^{\rm (em)}_{3d}|^2$. This suggests that comparing the intensities of different \ac{RIXS} channels  provides access to the  character of the involved \acp{MO}. This has been demonstrated in References~\citenum{Suljoti_ACIE_2013} and \citenum{Engel_JPCB_2014} for Fe(CO)$_5$ and [Fe(CN)$_6]^{4-}$, respectively.
Needless to say that this reasoning applies to other absorption edges as well.

In Figure \ref{fig:chemical_bonds}a, exemplary \ac{RIXS} spectra for the  Fe $L_3$-edge and the O $K_\alpha$-edge of Fe(CO)$_5$ are shown. In case of the Fe $L_3$-edge, there are, on the one hand side, local $d-d$ type radiative relaxation transitions of $n3d(e',e''$) to $2p$ character (see density difference plots in panel (b)). On the other hand side, inelastic peaks of $\sigma 3d(a_1'$) to $2p$ character are clearly discernible for different excitation energies. From the density difference plots (note especially the negative part shown in blue) in Figure \ref{fig:chemical_bonds}b the ligand to metal charge-transfer character of these core-hole refill transitions becomes apparent. Based on the simple model outlined before the intensity ratio of these two types of transitions can be used to characterize the degree of orbital mixing.
The same argument  applied to the O $K$-edge leads to the identification of the $n3d(e',e''$) to 1s transition as marker bands for the metal-to-ligand charge-transfer character. 
Taking  both results together gives a means for quantifying the \ac{MO} composition in terms of atomic orbitals or equivalently $\sigma$-donor/$\pi$-acceptor character of the metal-ligand bonding. In order to establish the general nature of this argument, the [Fe(CN)$_6]^{4-}$ complex has been investigated using \ac{RIXS} spectroscopy \cite{Engel_JPCB_2014}. Comparing band intensities between [Fe(CN)$_6]^{4-}$ and Fe(CO)$_5$ it was verified that CN$^-$ is a stronger $\sigma$-donor but weaker $\pi$ acceptor than CO, which is in accord with chemical intuition. 
%
\subsection{ Tracing catalytic activity }\label{sec:mn}
%
Following the building principles of natural water-oxidation systems, manganese-based nanostructured catalysts have attracted considerable attention.\cite{Wiechen_DT_2012,	Hocking_NC_2011, Singh_CM_2013}  X-ray spectroscopy is sensitive to the oxidation state of metal atoms. In Reference~\citenum{Bokarev_JPCC_2015}, this has been shown in a spectroscopic investigation of the redox evolution of	 manganese systems relevant for water photo-electro-oxidation.~\cite{khan14_18199,xiao15_872} In the experiment, Mn precursors with different oxidation states  have been doped into a nafion matrix, where all converted into Mn$^{\rm II}$. These species can be oxidized upon application of a bias and reduced under visible light illumination. To follow this reaction, \ac{XAS} as well as \ac{RIXS} spectra have been calculated for the model compounds $[{\rm Mn}^{\rm II}({\rm H}_2{\rm O})_6]^{2+}$, $[{\rm Mn}^{\rm III}({\rm H}_2{\rm O})_6]^{3+}$, and $[{\rm Mn}^{\rm IV}{\rm L}({\rm OMe})_3]^{+}$ (cf. Figure \ref{fig:mn}) \cite{Bokarev_JPCC_2015}. Although the actual experimental systems are much more complex, the local nature of the soft X-ray probe enables comparison with these model complexes. It turned out that for the present systems \ac{XAS} is much more sensitive to the oxidation state than \ac{RIXS}, hence only \ac{XAS} spectra are shown in Figure \ref{fig:mn}. Comparison between experiment and simulation is hampered by the fact, that the former are done for mixtures of species having different oxidation states. Nevertheless, both spectra not only show strong similarities, but also a distinct dependence on the oxidation state.

The \ac{XAS} spectra show the typical splitting between the $L_2$ and $L_3$ bands which is about 11 eV, almost independent  on the oxidation state. It follows that the SOC constant is about 7.3~eV. This large SOC constant leads to considerable coupling of core excited states of different spin, i.e. if $S$ denotes the spin of the electronic ground state, core-excited states are mixtures of $S-1$, $S$, and $S+1$ spin states. Any interpretation of the \ac{XAS} spectra has to account for this fact. The decomposition of the spin-orbit coupled wave functions into $\Delta S=0, \pm 1$ contributions is given in Figure \ref{fig:mn}b for the different species. 
\rt{Whereas the valence excited states show little mixing, i.e. they are either of $\Delta S=0$ or $\Delta S=-1$ character, the situation is considerably more complex for core-excited states, with the relative stability of spin-states being strongly affected by the core hole; for a more detailed discussion of the dependence on the actual species, see Reference  \citenum{Bokarev_JPCC_2015}.}

The XAS spectra can be assigned as follows (in the following we discuss the $L_3$ band  only, the $L_2$ band has a similar assignment, although the spin mixing is even stronger). (i) Mn$^{\rm II}$: Here, the low-energy shoulder (a) is of dominantly $2p \rightarrow 3d(t_{2g})$ character, comprising $\Delta S=0$ transitions. The main band (b) is shaped by intense $2p \rightarrow 3d(e_{g})$ transitions with $\Delta S=0, -1$. The peaks denoted by (c) are due to mostly spin-forbidden $\Delta S=-1$ transitions of mixed $t_{2g}/e_{g}$ character. (ii)  Mn$^{\rm III}$: The spectrum is dominated by  $2p \rightarrow 3d(e_{g})$ transitions, but also contains shake-up $3d(e_{g}) \rightarrow 3d(t_{2g})$ transitions in region (b).  Starting from the main peak (b)  $\Delta S=0, -1$ spin-mixing starts to play a role until the essentially spin-forbidden  $\Delta S=-1$ region (c). (iii) Mn$^{\rm IV}$: the spectrum is similar to Mn$^{\rm III}$, but also contains notable  contribution from $\Delta S=+1$ transitions in regions (b,d,e). 
The influence of the PT2 correction to the \ac{RASSCF} is also illustrated in Figure~\ref{fig:mn}.
Apart from an overstabilization of the ground state which needs to be compensated by a constant shift of the spectrum, its shape hardly changes as can be seen in Figure \ref{fig:mn} what is an evidence of mostly ionic metal-ligand bond character.


Analysis of  inelastic \ac{RIXS} spectra revealed that for all species most features are due to spin-forbidden transitions, mediated by core-excited states~\cite{Bokarev_JPCC_2015} similar to what has been discussed for iron complex in Section~\ref{sec:el_struct}.
Note that intricate spin mixing of the core-excited state gives rise to the exciting ultrafast spin-flip dynamics which can be triggered by \ac{XFEL} and \ac{HHG} X-ray pulses \cite{Wang_PRL_2017, Wang_MP_2017, Wang_PRA_2018}.

\subsection{ Aggregation and the challenge of multiple metal centers}
%
Straightforward extension of the \ac{RASSCF}-based approach to the calculation of core-excited states of multi-center metal compounds  would require active spaces that go beyond the capabilities of current hard- and software. Even if one would be able to perform such a calculation, the spectra would consist of hundreds of thousands of transitions, which would render any detailed interpretation of such spectra a challenging task. In order to address this problem, we have recently developed an approach which follows the logic of the so-called Frenkel exciton model.~\cite{schroter15_1,Preusse_SD_2016} The latter is used to describe the coupling of low-lying valence excitations in molecular aggregates and crystals, i.e. in situations where monomers assemble as a consequence of the van der Waals interaction. Since all what is needed is a localized excitation, the Frenkel exciton picture might apply to the present case of core-holes  even if   multiple metal centers in covalently bound complexes are considered.

The principal idea is to introduce metal center specific excitations which are in turn coupled by the Coulomb interaction. Thus, the Frenkel exciton Hamiltonian reads
\begin{eqnarray}
  \label{eq:hmat1}
  {\hat H}^{\rm FE} &=& \sum_M \sum_{A} E_{A_M} |A_M\rangle \langle A_M | \nonumber\\
              &+& \frac{1}{2} \sum_{MN}
                  \sum_{A,B,C,D}J_{MN}(A_MB_N,C_ND_M) \nonumber\\
&\times&  |A_M \rangle \langle D_M | \otimes |B_N \rangle \langle C_N  |\,.
\end{eqnarray}
where \bt{the states $A_M-D_M$ are eigenfunctions of the monomeric Hamiltonian, i.e. } $H_M|A_M\rangle=E_{A_M}|A_M \rangle$,  
and the Coulomb coupling can be expressed, e.g., in dipole approximation
\begin{eqnarray}
  \label{eq:jdip}
  J_{MN}(A_MB_N,C_ND_M) &\approx &\frac{\mathbf{d}_{A_MD_M}\cdot \mathbf{d}_{B_NC_N}}{|\mathbf{X}_{MN}|^3}\nonumber\\
                        & -& 3 \frac{(\mathbf{X}_{MN}\cdot\mathbf{d}_{A_MD_M})(\mathbf{X}_{MN}\cdot \mathbf{d}_{B_NC_N})}{|\mathbf{X}_{MN}|^5}\, ,
\end{eqnarray}
where $\mathbf{d}_{A_MD_M}= \langle A_M |\hat{{\mathbf d}}  |D_M \rangle$ is  the transition dipole matrix element and the vector ${ {\mathbf X}_{MN}}$ connects monomers $M$ and $N$.

\begin{figure}[tbh]
	\centering
	\includegraphics[width=0.8\textwidth]{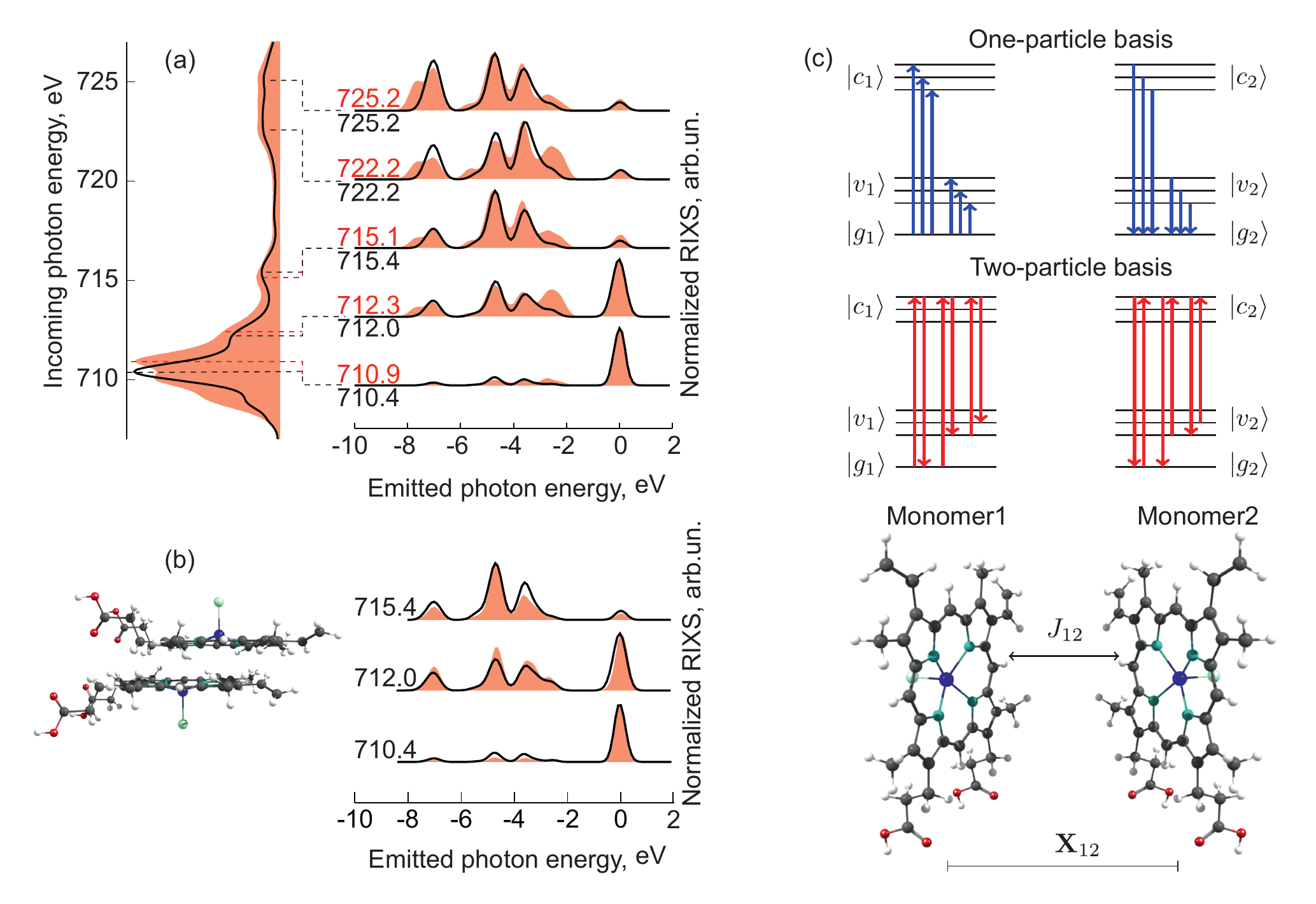}
	\caption{\label{fig:aggregates}
		 (a) Left: calculated normalized XAS for [heme B-Cl]$^0$ (black) and [heme B-H$_2$O]$^+$ (red filled curves). Right: normalized \ac{RIXS} spectra for selected excitation energies. 
		 (b) Calculated \ac{RIXS} spectra for the hemin dimer (red filled curves)  with the -COOH groups pointing in the same direction. The monomer spectrum is also shown ($\times 2$, black lines). Note that monomer and dimer XAS are indistinguishable on this scale.
		 (c) Different choices of excitonic bases and the corresponding transitions included in the dimer coupling. For the one-particle basis, only transitions from or to the ground state have been included. For the two-particle basis, de-excitations from an arbitrary core-excited state to any other state are allowed.  (adapted from Reference~\citenum{Preusse_SD_2016})}
\end{figure}

The approach is illustrated in Figure \ref{fig:aggregates}c for the example of a dimer. The different monomeric state manifolds will be denoted as ground $|g_M\rangle$, valence-excited, $|v_M\rangle$, and core-excited, $|c_M \rangle$, states.   Equation \eqref{eq:jdip} contains couplings between all possible transitions in the dimer
system. In the following, we will make use of the fact that the actual observables of interest,
namely \ac{XAS} and \ac{RIXS} spectral amplitudes, are of first and second order with
respect to the interaction with the external field. This suggests to employ either a one- or two-particle
basis as shown in Figure \ref{fig:aggregates}c. In the former, states of the type $|a_1g_2\rangle$ and $|g_1a_2\rangle$ ($a=v,c$) are incorporated, while the latter includes in addition states of type $|a_1,b_2\rangle$ ($a,b=v,c$) and is in principle exact for the dimer. Note that this effectively corresponds to a CISD-like treatment of the interaction between monomers in the composite system with X-ray specific preselection of configurations, while monomers are treated within the usual \ac{RASSCF} procedure. In order to make this approach computationally feasible, a number of approximations can be introduced as detailed in Reference~\citenum{Preusse_SD_2016}. Most important in this respect is the preselection of core-excited states to construct the monomeric basis functions, which is based on an energy window criterion for the excitation light pulse.

In a proof-of-concept study, this approach has been applied to investigate the signatures of aggregation of hemin in solution.~\cite{Preusse_SD_2016} Hemin shows a solvent dependent aggregation behavior, staying monomeric in polar solvents like ethanol or DMSO, while it forms dimers in water solution. Experimentally, the effect of aggregation was  addressed by means of soft X-ray Fe $L$-edge \ac{XAS} as well as by NXES and \ac{RIXS} on the example of DMSO and aqueous solutions.~\cite{atak14_9938,golnak15_3058}  The general shape of the spectra for both cases was quite similar and the pronounced difference in broadenings for \ac{RIXS} as well as a 1.3~eV energy shift in off-resonant  NXES was attributed to aggregation. However, judging the effect of aggregation one should account for the labile equilibrium between different species in solution that also could be a source of spectral changes.

In Figure~\ref{fig:aggregates}a, XAS  as well as selected \ac{RIXS} spectra are shown for the bare hemin molecule ([Heme\ B-Cl]$^0$) as well as for the case where the axial Cl$^-$ ligand has been exchanged by a water molecule ([Heme\ B-H$_2$O]$^+$). The \ac{XAS} spectrum shows a distinct sensitivity of the $L_3$ band to the type of  ligand. In the present case, changing the ligand from Cl$^-$ to H$_2$O causes a splitting of the $L_3$ peak in two components at 710.0\,eV and 710.9\,eV. While the main peak is due to $2p \rightarrow {d}_{{x^2-y^2}}$ and $d_{{z}^2}$ excitations, the splitting results from  the energetic lowering of $2p \rightarrow {d}_{{z^2}}$ with respect to $2p \rightarrow {d}_{{x^2-y^2}}$ transitions.

The RIXS spectra of the monomer show a  trend similar to the one  discussed in Section~\ref{sec:el_struct}.
For excitation energies
above 720\,eV, the core-excited states are mostly of quartet type. Here, the most intense emission goes to   valence-excited states with high quartet contributions. 
The RIXS spectra for both species  differ mostly in the inelastic peaks, whereas the elastic peak has a comparable intensity in both spectra.

The comparison of [Heme\ B-Cl]$^0$ dimer and monomer spectra is shown in Figure~\ref{fig:aggregates}b. The XAS spectra are essentially identical, i.e. there is almost no effect due to dimerization. This can be attributed to the intensities of the metal \(2p \rightarrow 3d\) and  \(3d\rightarrow 3d\) transitions relevant for $L$-edge X-ray spectra that are lower (due to smaller radial overlap and dipole selection rules) than those of the \(\pi \rightarrow \pi^*\) and \(n\rightarrow \pi^*\) transitions usually discussed in the case of aggregates of organic dyes.
However, distinct fingerprints of dimerization can be seen in RIXS, in both the inelastic and elastic features. Overall, the magnitude of the effect of dimerization is comparable to the differences observed upon ligand exchange in the monomer. Hence, the unequivocal identification of spectral features due to solvent dependent dimerization is not a straightforward task.

\section{CONCLUSIONS}
%
There is a \rt{plethora} of methods which have been developed to calculate core-excited states for prediction of various X-ray spectroscopic signals.
For \ac{TM} compounds beyond the $K$-edge, multi-reference approaches provide the methods of choice as they can deal with the intricate multiplet structure shaped by static and dynamic correlations and substantial SOC. However, limitations are due to the methods' complexity  and associated computational effort. Here, active space approaches and in particular RASSCF are playing a prominent role as they are, in principle, exact  and systematically improvable by increasing the size of the active space. 
In practice, however, methods like RASSCF cannot be considered as a ``black box'' due to the strong influence of the choice of the \ac{AS}. 
Recent progress in the field of \ac{DMRG} and automated \ac{AS} selection certainly mitigates this problem.~\cite{Stein_JCTC_2016a} For polymetallic complexes, merging established ideas of electronic structure theory with models developed in other areas such as the Frenkel exciton Hamiltonian could broaden the range of systems for theoretical X-ray spectroscopic studies.

To date, there is a considerable record of theoretical and experimental applications of frequency-domain X-ray techniques to study \ac{TM} complexes in various environments. With the emerging explicit time-domain techniques, either using \ac{XFEL} or \ac{HHG} facilities,  nonlinear spectroscopy in the X-ray domain will play a role for unraveling electron dynamics, comparable to that of UV/Vis in case of  (non-)Born-Oppenheimer electron-nuclear dynamics.~\cite{Zhang_TCC_2015}
\section*{ ACKNOWLEDGEMENTS}
We would like to acknowledge the financial support from the Deutsche Forschungsgemeinschaft via grant No. BO 4915/1-1 (S.I.B.) and Ku952/10-1 (O.K.) and the Deanship of Scientific Research (DSR), King Abdulaziz University, Jeddah, (Grant No. D-003-435).



\end{document}